\documentclass[letterpaper,12pt]{JHEP3}
\usepackage{array}
\usepackage{amsmath}
\usepackage{amssymb}

\newcommand{\eq}{\begin{equation}}
\newcommand{\en}{\end{equation}}
\newcommand{\eqn}{\begin{eqnarray}}
\newcommand{\enn}{\end{eqnarray}}

\newcommand{\beq}{\begin{equation}}
\newcommand{\eeq}{\end{equation}}
\newcommand{\M}{\ensuremath{\mathcal{M}}}

\newcommand{\tn}{\ensuremath{\tilde{n}}}

\newcommand{\tx}{\ensuremath{\tilde{x}}}
\newcommand{\ty}{\ensuremath{\tilde{y}}}
\newcommand{\ti}{\ensuremath{\tilde{I}}}
\newcommand{\tj}{\ensuremath{\tilde{J}}}
\newcommand{\tk}{\ensuremath{\tilde{K}}}

\newcommand{\hmu}{\ensuremath{\hat{\mu}}}
\newcommand{\hnu}{\ensuremath{\hat{\nu}}}
\newcommand{\hrho}{\ensuremath{\hat{\rho}}}
\newcommand{\hsigma}{\ensuremath{\hat{\sigma}}}
\newcommand{\hlambda}{\ensuremath{\hat{\lambda}}}
\newcommand{\he}{\ensuremath{{\hat{e}}}}
\newcommand{\hA}{\ensuremath{{\hat{A}}}}

\newcommand{\hB}{\ensuremath{{\hat{B}}}}
\newcommand{\hcD}{\ensuremath{{\hat{\mathcal{D}}}}}
\newcommand{\tI}{\ensuremath{\tilde{I}}}
\newcommand{\tJ}{\ensuremath{\tilde{J}}}
\newcommand{\tK}{\ensuremath{\tilde{K}}}

\title{The R-map and the Coupling of $\mathcal{N}=2$ Tensor Multiplets in 5 and 4
Dimensions}

\author{Murat G\"{u}naydin$^{1,2}$ , Sean McReynolds$^{1}$ and Marco Zagermann$^{3}$
 \\ 1. Physics Department,  Pennsylvania State University\\ University Park, PA 16802,
 USA \\
2. Kavli Institute of Theoretical Physics, University of California
\\ Santa Barbara, CA 93106 , USA
\\3.  Max-Planck-Institut f\"{u}r Physik,  F\"{o}hringer Ring 6\\
D-80805 Munich,   Germany\\ E-mail: \email{murat@phys.psu.edu,
sean@phys.psu.edu, zagerman@mppmu.mpg.de}}
\abstract{ We study the dimensional reduction of 5D, $\mathcal{N}=2$
Yang-Mills-Einstein supergravity theories (YMESGT) coupled to tensor
multiplets.
The resulting  4D theories  involve first order
interactions among tensor and vector fields  with mass terms.
If the 5D gauge group, $K$,  does not mix the 5D  tensor and vector fields,
the
 4D  tensor fields can be integrated out in favor of the 4D  vector fields
 and the resulting theory  is dual to  a  standard 4D  YMESGT (Integrating out the vector fields in favor of tensor fields instead seems to
 require nonlocal field redefinitions).
The  gauge group  has a block diagonal symplectic embedding and
is a semi-direct product of the 5D gauge group $K$ with a Heisenberg
group $\mathcal{H}^{n_T+1}$ of dimension $n_T+1$, where $n_T$ is the
number of tensor fields in five dimensions. There exists an infinite
family of theories, thus obtained, whose gauge groups are pp-wave
contractions of the simple noncompact  groups of type $SO^*(2N)$.
If, on the other hand, the 5D gauge group does mix the 5D  tensor and vector fields,
 the resulting 4D theory
is dual to a 4D YMESGT whose gauge group does, in general,
\emph{not} have a block diagonal symplectic embedding and  involves
additional topological terms.  The scalar potentials of the
dimensionally reduced theories studied in this paper naturally have
some of the ingredients that were found necessary for stable de
Sitter ground states in earlier studies. We comment on the relation
between the known 5D and 4D, $\mathcal{N}=2$ supergravities with
stable de Sitter ground states. } \preprint{\hepth{9999999}\\
NSF-KITP-05-90}
\begin{document}



\section{Introduction}
Four-dimensional supergravity theories with massive
antisymmetric tensor  fields \footnote{For some earlier work, see also
\cite{Tearlier}.}  have recently received
 a lot of attention \cite{T42,T41,T43} due to their relevance for
 string compactifications with background fluxes \cite{fluxes}
    or Scherk-Schwarz generalized dimensional reduction \cite{SS}.

In conventional string compactifications \emph{without} background fluxes
or geometric twists,
 \emph{massless}   two-forms  in the effective 4D theory naturally
  descend from the various types of p-form fields
    in     the 10D or 11D actions of string or M-theory.
A massless two-form in 4D is Hodge dual to a massless scalar field,
and upon such a dualization, the 4D effective theory is readily expressed
in terms of scalar fields and vector fields only (plus the gravitational sector
and the fermions). In an $\mathcal{N}=2$ compactification,  the resulting theories
then describe the coupling of massless vector and hypermultiplets to supergravity
without gauge interactions.

When fluxes or  geometric twists are switched on, however, the low
energy effective theories typically contain gauge interactions and
mass deformations, which in turn entail non-trivial scalar
potentials \footnote{Gauge interactions and non-trivial scalar potentials
can also occur when the scalar manifold exhibits certain types of singularities
 or is close to other special points in the moduli space, corresponding, e.g. to self-dual radii of circles etc. These gauge interactions and potentials
are often associated with additional light states, which, in the case of singularities,
 are typically localized at those singularities as e.g. in  \cite{Strominger}
  (for a  complete treatment of a concrete example in the language of
  gauged supergravity     see also   \cite{MZ01}).}.
In the presence of mass deformations for two-form
fields, the massive  two-form can no longer be directly dualized to
a scalar field. Instead, a massive two-form is dual to a massive
vector field \cite{TV}, and the relation to the standard formulation
of 4D gauged supergravity   in terms of scalar fields and vector
fields \cite{S4,dWLVP} is,  a priori,  less clear. In the  well understood
cases,  this relation involves the gauging of axionic isometries on
the scalar manifold, upon which the axionic scalar field can be
``eaten'' by a vector field     to render it massive
\cite{T42,T41,T43}.

In the context  of 4D, $\mathcal{N}=2$  supergravity, such mass
deformations have been primarily studied for two-forms that, before
the deformation, arise from dualizations of scalars of  the
quaternionic K\"{a}hler manifold of the hypermultiplet sector
\cite{T42}. It was only very  recently that such  mass deformations
were also studied  for tensor fields that are dual to scalars of the
special K\"{a}hler manifold \cite{dWST}.

Massive tensor fields also   play an important r\^{o}le in 5D, $\mathcal{N}=2$ gauged supergravity  \cite{GZ99}.
 In five dimensions, massless tensor fields are dual to massless vector fields when they are not charged with respect to any local gauge symmetry. In ungauged supergravity, two-form fields are therefore usually replaced by vector fields
\cite{GST1}. When gauge interactions are turned on, however,
the equivalence between two-form fields and vector fields is typically lost, and
one has to distinguish between them more carefully \cite{TPvN,GRW,PPvN85,Romans,GZ99,DHZ}.   In particular, two-form fields that transform non-trivially under some gauge group are possible, and
such two-forms are no longer equivalent to vector fields \footnote{A reformulation of 5D, $\mathcal{N}=8$ gauged supergravity which treats vector and tensor fields more symmetrically has recently been given in \cite{new}.}. This can be understood from the fact that the charged tensor fields acquire a mass, and massive tensors in 5D have a different number of degrees of freedom than vectors.
In the conventional formulation of such 5D gauged supergravity theories
 with tensor fields,  the tensor fields $B_{\mu\nu}^M$    enter the Lagrangian via
 first order terms of the form \cite{GRW,PPvN85,Romans,GZ99,DHZ}
 \begin{equation}
 \Omega_{MN} B^M \wedge \mathcal{D} B^{N}
 \end{equation}
where $\Omega_{MN}$ is a symplectic metric,   $\mathcal{D}B^N = dB^N +g \Lambda_{IM}^{N} A^{I} \wedge B^{M}$,
and $\Lambda_{IM}^{N}$ denotes the transformation matrix of the tensor fields with
 respect to the gauge group gauged by the vector fields $A_{\mu}^{I}$
 and with gauge coupling $g$.
Reiterating the resulting field equations, half of the tensors can be eliminated, and one obtains    second order field equations for
 the remaining ones with mass terms due to a $B^{M} \wedge \ast B^{N}$ coupling
 in the Lagrangian.

Whereas the dimensional reduction of  \emph{ungauged} 5D
supergravity to 4D has been studied quite extensively in
 the literature, surprisingly little
is known about the dimensional reduction of 5D \emph{gauged}
supergravity \emph{with tensor fields}.
For example, the
$\mathcal{N}=8$ AdS graviton supermultiplet involves both vector and
tensor fields in five dimensions \cite{GM}.
  Hence gauging the
maximal supergravity in five dimensions requires that some of the
vector fields of the ungauged theory be dualized to tensor
multiplets \cite{GRW,PPvN85}.  Remarkably, the $SU(3,1)$ gauged
$\mathcal{N}=8$ supergravity constructed in \cite{GRW2} has
 a stable ground state that preserves two supersymmetries and has a vanishing cosmological constant.
 The general properties of the compactification of the $SU(3,1)$
gauged  5D, $\mathcal{N}=8$
 supergravity  down to four dimensions
were originally investigated in \cite{GRW2}. More recently, a more
detailed analysis of the dimensional reduction of 5D  gauged
$\mathcal{N}=8$ supergravity down to four dimensions was given by
Hull \cite{Hull}, but to the best of our knowledge, a complete
analysis, in particular for $\mathcal{N}=2$, has never been given.
As the  naive dimensional reduction is expected to involve massive
two-forms of some sort, it is important to close this gap in the
literature and to compare the result with the current work   on 4D
massive two-forms \cite{T41,T42,T43,new} and the standard formulation of
gauged supergravity theories in 4D \cite{S4,dWLVP}.

As the resulting theory only involves    the (very) special
K\"{a}hler gemetry of the vector multiplet sector in 4D, and since
the tensors are expected to transform nontrivially  under (in general
  non-Abelian) gauge symmetries, the resulting theories are
expected to be  different from the ones studied in the recent works
\cite{T42} on hypermultiplet scalars.

The   dimensional reduction of 5D, $\mathcal{N}=2$ gauged supergravity with tensor multiplets to 4D could also be interesting for the recent attempts
to find
stable de Sitter ground states in extended supergravity theories \cite{GZ2,dS,4dS,CS}.
So far, the only known examples for such stable de Sitter vacua   were found in 5D, $\mathcal{N}=2$ gauged  supergravity theories   with tensor fields \cite{GZ2,CS} and in
certain 4D, $\mathcal{N}=2$ gauged supergravity theories \cite{4dS}.
As for the latter type of theories, the authors of \cite{4dS}  identified a number of ingredients that were necessary to obtain stable de Sitter vacua.
These include non-Abelian non-compact gauge groups,  de Roo-Wagemans rotation angles \cite{dRW}      and gaugings of subgroups of the R-symmetry group.
Interestingly, gaugings of the R-symmetry group also play a r\^{o}le
for the known   5D theories with stable de Sitter vacua \cite{GZ2,CS}. Also, the known 5D examples involve non-compact gauge groups. However, in 5D, these groups can be Abelian and still give rise to stable de Sitter vacua. Furthermore, the known 5D
models involve charged tensor multiplets, whereas
 de Roo-Wagemans rotation
angles    are not well-defined in 5D. One of the important results
of our paper is that the dimensional reduction of 5D,
$\mathcal{N}=2$ gauged supergravity with tensor multiplets to 4D
always leads to non-Abelian non-compact gauge groups, no matter what
the 5D gauge group is. Furthermore, one always introduces something similar to de Roo-Wagemans rotation angles    in this reduction process. We do  not
consider gaugings of the R-symmetry group in this paper, but putting
the above together, one might wonder whether  the dimensionally
reduced 5D theories with tensor fields could  give rise to 4D stable de
Sitter vacua, perhaps after switching on R-symmetry gaugings and/or
suitable truncations or extensions.

Motivated by these and other possible  applications, we will, in this paper,
systematically study  the dimensional reduction of 5D, $\mathcal{N}=2$ gauged supergravity with tensor multiplets to 4D.

The outline of this paper is as follows. In Section 2, we briefly
recapitulate the structure of  5D, $\mathcal{N}=2$ ungauged
Maxwell-Einstein supergravity theories  (MESGTs). In Section 3, we
review the gaugings of these theories which require the introduction
of tensor fields. Here, two cases are to be distinguished: (i) The
vector fields of the ungauged theory transform in a completely
reducible representation of the prospective gauge group, or (ii)
they form a representation that is reducible, but not
\emph{completely} reducible \cite{5Dgroup}. In Section 4, we
dimensionally reduce the theories of type (i) to 4D. Section 5
discusses the r\^{o}le played by the massive  two-forms and vector
fields in the resulting 4D theories. The dimensionally reduced
theories have a first order interaction between  two-form and vector
fields that is reminiscent of the Freedman-Townsend model \cite{FT}
and looks like a concrete
 realization of the formalism of  \cite{dWST}.
We then eliminate the tensor
fields in favor of   vector fields, which are indeed massive. The
opposite elimination of the vector fields in terms of the tensor
fields  meets some difficulties and might be possible only in a
rather  non-trivial way. In Section 6, we show that, after suitable
symplectic rotations, the resulting theory without the two-forms can
be mapped to a standard gauged supergravity theory in 4D in which the gauge group has a block diagonal  symplectic embedding. This
theory has a gauge group of the form
  ($ K \ltimes \mathcal{H}^{n_{T}+1} ) $, which
is the semidirect product of the 5D gauge group $K$  with the $(n_{T}+1)$-dimensional
Heisenberg group $\mathcal{H}^{n_{T}+1}$ generated by $n_T$ translation
generators and a central charge ($n_{T}$ denotes the number of
tensor multiplets in five dimensions, which is always even).
The case (ii) of not completely recducible representations is
briefly sketched in Section 6.3.
The dimensional reduction of theories with completely reducible representations
in 5D
parallels the situation in the $\mathcal{N}=8$ theory described by
Hull in \cite{Hull}, as explained in Section 7, where we also
comment on the relation to the ``unified'' supergravity theories
studied in \cite{GMcRZ1}.  In section 8, we study some properties of
the scalar potential and comment on the relation to extended supergravity theories with stable de Sitter ground states.   Appendix A, finally, contains some
details of the dimensional reduction.



\section{5D, $\mathcal{N}=2$ Maxwell-Einstein supergravity theories}
Five-dimensional minimal supergravity can be coupled to vector, tensor and hypermultiplets \cite{GST1,GST2,Sierra,GZ99,CD,GZ3,5Dgroup}.
 Hypermultiplets are irrelevant for this paper and will henceforth be ignored.
In five dimensions, massless  uncharged
  vector fields and
    massless
 uncharged  two-form fields are dual to one another.
 At the level of ungauged supergravity theories, the
distinction between vector and tensor multiplets is therefore unnecessary,
and one can, without loss of generality,
dualize all tensor fields to vector fields.  These theories
are often referred to as ``Maxwell-Einstein supergravity theories'' (``MESGTs'')  and were first constructed in \cite{GST1}. Our notation in this paper
follows that of \cite{GST1,GZ99}, except that we will
 put a hat on all
five-dimensional spacetime and tangent space
indices,
 as well as on all fields that decompose nontrivially into four-dimensional fields, as will
  become obvious below.

 The 5D, $\mathcal{N}=2$    supergravity multiplet consists of
the f\"{u}nfbein $\hat{e}_{\hat{\mu}}^{\hat{m}}$, two gravitini
$\hat{\psi}_{\hat{\mu}}^{i}$ $(i=1,2)$ and one vector field $\hat{A}_{\hat{\mu}}$
(the ``graviphoton'').   A vector multiplet contains a vector field
$\hat{A}_{\hat{\mu}}$, two ``gaugini'' $\hat{\lambda}^{i}$ and one real scalar field, $\varphi$. Coupling $\tilde{n}$ vector multiplets to supergravity, the total bosonic
field content is then
\begin{displaymath}
\{ \hat{e}_{\hat{\mu}}^{\hat{m}}, \hat{A}_{\hat{\mu}}^{\tilde{I}}, \varphi^{\tilde{x}} \},
\end{displaymath}
where, as usual, the graviphoton and the $\tn$ vector fields from the $\tn$
vector multiplets have been combined  into one $(\tn+1)$-plet of vector fields
$\hat{A}_{\hat{\mu}}^{\tilde{I}}$ $(\tilde{I}=1,\ldots,\tn+1)$. The indices
$\tx, \ty,
\ldots$ denote the curved indices of the
 $\tn$-dimensional target manifold, $\mathcal{M}^{(5)}$,
of the scalar fields.

The bosonic part of the Lagrangian is given by (for the fermionic
part and further details, see \cite{GST1})
\begin{eqnarray}
 \mathcal{L}^{(5)} 
 &=&
 -\frac{1}{2} \hat{e}\hat{R}
-\frac{1}{4}\hat{e} {\stackrel{\circ}{a}}_{\tilde{I}\tilde{J}}
\hat{F}_{\hat{\mu}\hat{\nu}}^{\tilde{I}}\hat{F}^{\tilde{J}\hat{\mu}\hat{\nu}}-\frac{\hat{e}}{2}g_{\tx\ty}(\partial_{\hat{\mu}}\varphi^{\tx})
(\partial^{\hat{\mu}}\varphi^{\ty})\nonumber\\
&&+\frac{1}{6\sqrt{6}}C_{\tI\tJ\tK}\hat{\epsilon}^{\hat{\mu}\hat{\nu}\hat{\rho}\hat{\sigma}\hat{\lambda}}
\hat{F}_{\hat{\mu}\hat{\nu}}^{\tI}
{\hat{F}}_{\hat{\rho}\hat{\sigma}}^{\tJ}
\hA_{\hat{\lambda}}^{\tK}
\label{start1}
 \end{eqnarray}

where  $\hat{e}$ and $\hat{R}$
 denote, respectively,   the f\"{u}nfbein determinant and scalar
curvature of spacetime.  $\hat{F}_{\hat{\mu}\hat{\nu}}^{\ti}\equiv 2\partial_{[\hmu}\hat{A}_{\hnu]}^{\tI}$ are the standard
Abelian field strengths of the vector fields $\hat{A}_{\hat{\mu}}^{\ti}$. The
metric, $g_{\tx\ty}$, of the scalar manifold $\M^{(5)}$
 and the matrix ${\stackrel{\circ}{a}}_{\ti\tj}$ both depend
on the scalar fields $\varphi^{\tx}$. The completely symmetric
tensor $C_{\ti\tj\tk}$, by contrast, is constant.

The entire
$\mathcal{N}=2$ MESGT (including the fermionic terms and the
supersymmetry transformation laws that we have suppressed) is
uniquely determined by    $C_{\ti\tj\tk}$ \cite{GST1}. More
explicitly, $C_{\ti\tj\tk}$ defines  a cubic polynomial,
$\mathcal{V}(h)$,
 in $(\tn+1)$
real variables $h^{\ti}$ $(\ti=1,\ldots,\tn+1)$,
\begin{equation}
\mathcal{V}(h):=C_{\ti\tj\tk}h^{\ti}h^{\tj}h^{\tk}\ .
\end{equation}
This polynomial defines a metric, $a_{\ti\tj}$,
 in the (auxiliary) space $\mathbb{R}^{(\tn+1)}$ spanned by the $h^{\ti}$:
\begin{equation}\label{aij}
a_{\ti\tj}(h):=-\frac{1}{3}\frac{\partial}{\partial h^{\ti}}
\frac{\partial}{\partial h^{\tj}} \ln \mathcal{V}(h)\ .
\end{equation}
The  $\tn$-dimensional    target space, $\mathcal{M}^{(5)}$, of the scalar
fields $\varphi^{\tx}$ can then be represented as the hypersurface
\cite{GST1}
\begin{equation}\label{hyper1}
{\cal V} (h)=C_{\ti\tj\tk}h^{\ti}h^{\tj}h^{\tk}=1 \ ,
\end{equation}
with $g_{\tx\ty}$ being the pull-back of (\ref{aij}) to $\mathcal{M}^{(5)}$:
\begin{equation}
g_{\tx\ty}(\varphi)=\frac{3}{2} (\partial_{\tx} h^{\tilde{I}}) (\partial_{\ty} h^{\tilde{J}}) a_{\tilde{I} \tilde{J}}|_{\mathcal{V}=1} \ .   \label{kin}
\end{equation}
Finally, the quantity ${\stackrel{\circ}{a}}_{\ti\tj}(\varphi)$ appearing in
(\ref{start1}), is given by the componentwise restriction of
$a_{\ti\tj}$ to $\mathcal{M}^{(5)}$:
\[
{\stackrel{\circ}{a}}_{\ti\tj}(\varphi)=a_{\ti\tj}|_{{\cal V}=1} \ .
\]



\section{Charged tensor fields in five dimensions}
In the previous section, we considered 5D, $\mathcal{N}=2$ ungauged supergravity theories, in which all  potential
tensor fields     can be dualized to vector  fields, and the whole theory can be expressed in terms of vector fields
 only.
In the presence of gauge interactions, however,
  this
 equivalence between vector and tensor fields generally breaks down,
and one carefully has to distinguish between them \cite{GZ99}.

  In the Maxwell-Einstein supergravity theories of the previous section,
there are, in principle, the options for  two types of possible  gauge groups. One type
corresponds to the gauging of a subgroup of the R-symmetry group, $SU(2)_R$,
which acts on the index $i$ of the fermions. This type of gauging is irrelevant for the present analysis and will no longer
 be considered in this paper, except for a brief mentioning in Section 8.
The other type of gauging correspond to gaugings of symmetries of the tensor $C_{\ti\tj\tk}$. As $C_{\ti\tj\tk}$ determines the entire supergravity theory, such symmetries, if they exist,
 are automatically symmetries of the whole Lagrangian, and in particular, they are isometries of the scalar manifold $\mathcal{M}^{(5)}$. We denote by $G$ the group of
linear transformations of the $h^{\ti}$  and $\hat{A}_{\hat{\mu}}^{\ti}$
that leave the tensor $C_{\ti\tj\tk}$ invariant. They are generated by
infinitesimal transformations of the form
\begin{equation} \label{hAtrafo}
h^{\ti}\rightarrow {M_{(r)}^{\ti}}_{\tj}h^{\tj}, \quad \hat{A}_{\hat{\mu}}^{\ti}\rightarrow
{M_{(r)}^{\ti}}_{\tj}\hat{A}_{\hmu}^{\tj}
\end{equation}
with
\begin{displaymath}
M_{(r)}^{\ti'}{}_{(\ti}C_{\tj\tk)\ti'} =0 \ .
\end{displaymath}
Here, $r=1,\ldots,\dim(G)$ counts the generators of $G$.

In order to turn a  subgroup $K\subset   G$    into a \emph{local}
(i.e., Yang-Mills-type) gauge symmetry, the $(\tn+1)$-dimensional representation
of $G$ defined by the action (\ref{hAtrafo}) has to contain the adjoint representation of $K$ as a subrepresentation. If this is the case, there are two possibilities:\\
(i) The decomposition of the $(\tn+1)$-dimensional representation of $G$
    with respect to $K$ is completely reducible.\\
(ii)
 The decomposition of the $(\tn+1)$-dimensional representation of $G$
    with respect to $K$ is  reducible, but \emph{not} completely reducible.\\

Case (i), which is always the case for all connected
 semisimple and for all compact gauge groups,
 was analyzed in \cite{GZ99}. The second possibility (ii) has been later
 studied  in  \cite{5Dgroup}. We will first consider the first case (i), and later comment on the second case in Section 6.3.

If the $(\tn+1)$-dimensional representation of $G$ is completely reducible,
the vector fields $\hat{A}_{\hmu}^{\tI}$ decompose into a direct sum of
vector fields
$\hA_{\hmu}^I$  $(I=1,\ldots, n_{V}= \dim{K})$ in the adjoint  of $K\subset G$
plus possible additional non-singlets $\hA_{\hmu}^{M}$ $(M=1,\ldots, n_{T}=(\tn+1-n_{V}))$ of $K$.\footnote{If there are also singlets of $K$
in the $\mathbf{(\tn + 1 ) }$ of $G$, we include them in the set of
vector fields in the adjoint of (an appropriately enlarged)  $K$, where they
 simply correspond to Abelian factors under which nothing is charged.}  In order for the gauging of $K$ to be possible, the non-singlet vectors
$\hA_{\hmu}^{M}$ have
  to be converted to antisymmetric tensor fields $\hB_{\hmu\hnu}^{M}$ prior to the gauging \cite{GZ99}. We denote by $f_{IJ}^{K}$ the structure constants of  the gauge group  $K\subset G$ and use $\Lambda_{IM}^{N}$
    for the $K$-transformation matrices of the tensor fields $\hat{B}_{\hat{\mu}\hat{\nu}}^{M}$.
     The transformation matrices $\Lambda_{IM}^{N}$ of the tensor fields have to
     be symplectic with respect to an antisymmetric  metric $\Omega_{MN}$ :
     \begin{equation}
     \Lambda_{IM}^{N}\Omega_{NP}+\Lambda_{IP}^{N}\Omega_{MN}=0 \label{OL1}
     \end{equation}
and are related to the coefficients $C_{IMN}$ of the $C_{\tI\tJ\tK}$ tensor via
     \begin{equation}
     \Lambda_{IM}^{N}=\frac{2}{\sqrt{6}} \Omega^{NP} C_{IMP} \Longleftrightarrow
     C_{IMN}=\frac{\sqrt{6}}{2}\Omega_{MP}\Lambda_{IN}^{P}, \label{OL2}
     \end{equation}
where $\Omega_{MN}\Omega^{NP}=\delta_{M}^{P}$.

     The transformation matrices $M_{(I)\tK}^{\tJ}$ of eq. (\ref{hAtrafo})
     that correspond to the subgroup $K\subset G$ consequently decompose as follows
     \begin{equation}
     M_{(I)\tK}^{\tJ}=   \left( \begin{array}{cc}
     f_{IK}^{J} & 0 \\
     0 & \Lambda_{IM}^{N} \end{array} \right) . \label{MfB}
     \end{equation}
     Denoting the $K$ gauge coupling constant by $g$,
the Yang-Mills  field strengths $\hat{\mathcal{F}}_{\hmu\hnu}^I$ read
\begin{equation}
\hat{\mathcal{F}}_{\hat{\mu}\hat{\nu}}^{I} \equiv  2\partial_{[\hat{\mu}}\hat{A}_{\hat{\nu}]}^{I} + gf_{JK}^{I}\hat{A}_{\hat{\mu}}^{J}\hat{A}_{\hat{\nu}}^{K},
\end{equation}
  and the covariant derivatives of the tensor fields are defined by
  \begin{equation}\label{covB}
  \hat{\mathcal{D}}_{[\hat{\mu}}\hat{B}_{\hnu\hrho]}^M\equiv \partial_{[\hmu}
  \hat{B}_{\hnu\hrho]}^M +g\hat{A}_{[\hmu}^{I}\Lambda_{IN}^{M}\hat{B}_{\hnu\hrho]}^{N}.
  \end{equation}
  It is sometimes useful to combine the field strengths $\hat{\mathcal{F}}_{\hat{\mu}\hat{\nu}}^{I}$ and the tensor fields
  $\hat{B}_{\hmu\hnu}^M$ into  an $(\tn+1)$-plet of two-forms,
  \begin{equation}
  \hat{\mathcal{H}}_{\hat{\mu}\hat{\nu}}^{\tI}:=(\hat{\mathcal{F}}_{\hat{\mu}\hat{\nu}}^{I},\hat{B}_{\hmu\hnu}^M)
  \end{equation}
   Using the $K$-covariant derivative of the scalars given by
\begin{equation}
\hat{\mathcal{D}}_{\hmu}\varphi^{\tx}\equiv \partial_{\hmu}\varphi^{\tx}
+g\hat{A}_{\hmu}^{I}K_{I}^{\tx},
\end{equation}
where $K_{I}^{\tx}$ denotes the Killing vectors on $\mathcal{M}^{(5)}$ that correspond to $K\subset G$,
  the bosonic part of the Lagrangian then reads \cite{GZ99}
       \begin{eqnarray}
 \mathcal{L}^{(5)} 
 &=&
 -\frac{1}{2} \hat{e}\hat{R}
-\frac{1}{4}\hat{e} {\stackrel{\circ}{a}}_{\tilde{I}\tilde{J}} \hat{\mathcal{H}}_{\hat{\mu}\hat{\nu}}^{\tilde{I}}\hat{\mathcal{H}}^{\tilde{J}\hat{\mu}\hat{\nu}}-\frac{\hat{e}}{2}g_{\tx\ty}(\hat{\mathcal{D}}_{\hat{\mu}}\varphi^{\tx})
(\hat{\mathcal{D}}^{\hat{\mu}}\varphi^{\ty})\nonumber\\
&&+\frac{1}{6\sqrt{6}}C_{IJK}\hat{\epsilon}^{\hat{\mu}\hat{\nu}\hat{\rho}\hat{\sigma}\hat{\lambda}}\Big\{
\hat{F}_{\hat{\mu}\hat{\nu}}^{I}
{\hat{F}}_{\hat{\rho}\hat{\sigma}}^{J}
\hA_{\hat{\lambda}}^{K}+\frac{3}{2}g{\hat{F}}_{\hmu\hnu}^{I}\hA_{\hrho}^{J}
(f_{LF}^{K}\hA_{\hsigma}^{L}\hA_{\hlambda}^{F})\nonumber\\
&& \qquad \qquad \qquad \qquad   +\frac{3}{5}g^{2}(f_{GH}^{J}\hA_{\hnu}^G \hA_{\hrho}^{H})(f_{LF}^{K}\hA_{\hsigma}^{L}\hA_{\hlambda}^{F})\hA_{\hmu}^{I}
 \Big\} \nonumber\\
 && +\frac{1}{4g}\hat{\epsilon}^{\hmu\hnu\hrho\hsigma\hlambda}\Omega_{MN}\hB_{\hmu\hnu}^{M}\hcD_{\hrho}\hB_{\hsigma\hlambda}^{N} -\he g^2 P^{(T)} .    \label{start}
 \end{eqnarray}
Here, the scalar potential $P(T)$ is given by
\begin{equation}
P^{(T)}=\frac{9}{8}{\stackrel{\circ}{a}}_{MN}(\Lambda_{JP}^{M}h^{J}h^{P})(\Lambda_{IQ}^{N}h^{I}h^{Q}).
\end{equation}



\section{The dimensional reduction to four dimensions}
In this section, we dimensionally reduce the theories described in the
previous sections to four dimensions. For the sake of clarity, and to set up our notation, let us first recapitulate  the dimensional reduction of the \emph{ungauged} MESGTs without tensor fields  of section 2.


\subsection{The ungauged MESGTs and (very) special K\"{a}hler geometry}
The dimensional reduction of the bosonic sector of 5D, $\mathcal{N}=2$ MESGTs
to four dimensions was first carried out in \cite{GST1} and  further studied in \cite{dWVP1}.


\subsubsection{The reduced action}
We split the f\"{u}nfbein as follows
\begin{equation}
\he_{\hmu}^{\hat{m}} = \left(   \begin{array}{cc}e^{-\frac{\sigma}{2}}e_{\mu}^{m} & \hspace{2mm}  2W_{\mu}e^{\sigma}\\
e_{5}^{m}=0 & \hspace{2mm}  e^{\sigma} \end{array}  \right) ,\label{fuenfbein}
\end{equation}
which implies $\he=e^{-\sigma} e$, where $e=\det (e_{\mu}^{m})$.
The Abelian field strength of $W_{\mu}$ will be denoted by
$W_{\mu\nu}$:
\begin{equation}
W_{\mu\nu}\equiv 2\partial_{[\mu} W_{\nu]}.
\end{equation}
The vector fields $\hA_{\hmu}^{\tI}$ are decomposed into a 4D vector field, $A_{\mu}^{\tI}$, and a 4D scalar, $A^{\tI}$, via
\begin{equation}
\hA_{\hmu}^{\tI}=  \left( \begin{array}{c}
\hA_{\mu}^{\tI}\\
\hA_{5}^{\tI}
\end{array} \right) = \left( \begin{array}{c}
A_{\mu}^{\tI}+2W_{\mu}A^{\tI}\\
A^{\tilde{I}} \end{array}    \right). \label{vector}
\end{equation}
In the following, all 4D    Abelian   field strengths $F_{\mu\nu}^{\tI}$ refer to $A_{\mu}^{\tI} $, which is the invariant combination with respect to the $U(1)$ from the compactified circle:
\begin{equation}
F_{\mu\nu}^{\tI}\equiv2 \partial_{[\mu}A_{\nu]}^{\tI}.
\end{equation}
The dimensionally reduced action of the ungauged theory (i.e., of eq. (\ref{start1})) is then
\begin{eqnarray}
 e^{-1}\mathcal{L}^{(4)} &=&-\frac{1}{2}R
 -\frac{1}{2}e^{3\sigma}W_{\mu\nu}W^{\mu\nu}-\frac{3}{4}\partial_{\mu}\sigma
 \partial^{\mu}\sigma \nonumber\\
 &&-\frac{1}{4}e^{\sigma}{\stackrel{\circ}{a}}_{\tI\tJ}(F_{\mu\nu}^{\tI}+2W_{\mu\nu}A^{\tI}
 )( F^{\tJ\mu\nu}+2W^{\mu\nu}A^{\tJ})\nonumber\\
&& -\frac{1}{2}e^{-2\sigma}{\stackrel{\circ}{a}}_{\tI\tJ}(\partial_{\mu} A^{\tI})(\partial^{\mu}A^{\tJ}) -\frac{3}{4}
{\stackrel{\circ}{a}}_{\tI\tJ}(\partial_{\mu}h^{\tI})(\partial^{\mu}h^{\tJ})\nonumber\\
&&+\frac{e^{-1}}{2\sqrt{6}} C_{\tI\tJ\tK}\epsilon^{\mu\nu\rho\sigma}
\Big\{ F_{\mu\nu}^{\tI}F_{\rho\sigma}^{\tJ}A^{\tK} + 2
F_{\mu\nu}^{\tI}W_{\rho\sigma}A^{\tJ}A^{\tK} \nonumber \\
&& \qquad \qquad \qquad \qquad   +\frac{4}{3}W_{\mu\nu}W_{\rho\sigma} A^{\tI}A^{\tJ}A^{\tK} \Big\}
\label{redlag}
\end{eqnarray}


\subsubsection{(Very) special K\"{a}hler geometry}
This can be recast in the form of special K\"{a}hler geometry (in
fact, ``very special'' K\"{a}hler geometry in the terminology
introduced in \cite{dWVP2})   as follows \cite{GST1}. Define complex
coordinates
\begin{equation}
z^{\ti}:=\frac{1}{\sqrt{3}}A^{\tI} + \frac{i}{\sqrt{2}}\tilde{h}^{\tI}
\end{equation}
where
\begin{equation}
\tilde{h}^{\ti}:=e^\sigma h^{\ti}. \label{htildeI}
\end{equation}
These $(\tn+1)$ complex coordinates $z^{\ti}$ can be interpreted as the inhomogeneous coordinates
corresponding to  the $(\tn+2)$-dimensional complex vector
\begin{equation}
X^A=  \left( \begin{array}{c}
X^{0}\\
X^{\ti}\end{array}  \right) =    \left( \begin{array}{c}
1 \\
z^{\ti}
\end{array} \right)  .
\end{equation}
Introducing the ``prepotential''
\begin{equation}
F(X^A)=-\frac{\sqrt{2}}{3} C_{\ti\tj\tk}\frac{X^{\tI}X^{\tJ}X^{\tK}}{X^{0}} \label{prepot}
\end{equation}
and the symplectic section\footnote{One  should perhaps emphasize
that, fundamentally, the Lagrangian can be expressed in terms of
a symplectic section
$(X^A,F_A)$ without direct  reference to a prepotential.
In fact, a generic  symplectic section need not be such
that  $F_A =  \partial_A
F$ for some function $F$.
  However, one can always go to a symplectic basis where the new $F_{A}$
is, at least locally, the derivative of a prepotential $F$ (see,
e.g., \cite{CRTVP}).}

\begin{equation} \label{symsec1}
\left( \begin{array}{c}
X^A\\
F_{A}   \end{array}
\right)  \equiv  \left( \begin{array}{c}
X^A\\
 \partial_{A}F \end{array}
\right) \ ,
\end{equation}
one can define a K\"{a}hler potential
\begin{eqnarray}
K(X(z),\bar{X}(\bar{z}))&:=&-\ln [i\bar{X}^{A}F_{A}-iX^{A}\bar{F}_{A}] \label{symK}\\
&=&-\ln  \left[  i\frac{\sqrt{2}}{3}C_{\tI\tJ\tK}(z^{\tI}-\bar{z}^{\tI})(z^{\tJ}-\bar{z}^{\tJ})(z^{\tK}-\bar{z}^{\tK})  \right]
\end{eqnarray}
and a ``period matrix''
\begin{equation}
\mathcal{N}_{AB}:=\bar{F}_{AB}+2i\frac{\textrm{Im}(F_{AC})\textrm{Im}(F_{BD})X^{C}X^{D}}{\textrm{Im}(F_{CD})X^{C}X^{D}},
\end{equation}
where $F_{AB}\equiv \partial_{A}\partial_{B}F$ etc.
The particular (``very special'') form (\ref{prepot}) of the prepotential
 leads to
 \begin{equation}
g_{\tI\bar{\tJ}}\equiv \partial_{\tI}\partial_{\bar{\tJ}}K=\frac{3}{2} e^{-2\sigma} {\stackrel{\circ}{a}}_{\tI\tJ}
 \end{equation}
for the K\"{a}hler metric, $g_{\tI\bar{\tJ}}$, on the scalar manifold $\M^{(4)}$  of the four-dimensional theory, and
\begin{eqnarray}
\mathcal{N}_{00}&=&-\frac{2\sqrt{2}}{9\sqrt{3}}C_{\tI\tJ\tK}A^{\tI}A^{\tJ}A^{\tK}
-\frac{i}{3} \left( e^{\sigma}{\stackrel{\circ}{a}}_{\tI\tJ} A^{\tI}A^{\tJ}+\frac{1}{2} e^{3\sigma} \right)    \nonumber  \\
\mathcal{N}_{0\tI}&=&\frac{\sqrt{2}}{3}C_{\tI\tJ\tK}A^{\tJ}A^{\tK}+\frac{i}{\sqrt{3}} e^{\sigma} {\stackrel{\circ}{a}}_{\tI\tJ} A^{\tJ} \nonumber   \\
\mathcal{N}_{\tI\tJ}&=&  -\frac{2\sqrt{2}}{\sqrt{3}}C_{\tI\tJ\tK}A^{\tK}-ie^{\sigma}
{\stackrel{\circ}{a}}_{\tI\tJ} \label{NIJ}
\end{eqnarray}
for the period matrix $\mathcal{N}_{AB}$.
Defining
\begin{equation}
F_{\mu\nu}^{0}:=-2\sqrt{3}W_{\mu\nu}, \label{WF0}
\end{equation}
  the dimensionally reduced Lagrangian (\ref{redlag}) simplifies to
  \begin{eqnarray}
   e^{-1}\mathcal{L}^{(4)} &=&-\frac{1}{2}R  -
   g_{\ti\bar{\tJ}}    (\partial_{\mu}z^{\ti})(\partial^{\mu} \bar{z}^{\tJ})
   \nonumber \\
   &&+\frac{1}{4}\textrm{Im}(\mathcal{N}_{AB})F_{\mu\nu}^{A}F^{\mu\nu B}-\frac{e^{-1}}{8}
   \textrm{Re} (\mathcal{N}_{AB})\epsilon^{\mu\nu\rho\sigma}
   F_{\mu\nu}^{A}F_{\rho\sigma}^{B}.\label{redlag1b}
   \end{eqnarray}
In terms of    the selfdual and anti-selfdual field strengths,
\begin{eqnarray}
F_{\mu\nu}^{A\pm}&\equiv &\frac{1}{2} \left( F_{\mu\nu}^{A} \mp
\frac{i}{2} e\epsilon_{\mu\nu\rho\sigma}F^{A\rho\sigma}  \right)
\nonumber \\
F^{A\pm \mu\nu}&\equiv &\frac{1}{2} \left( F^{A \mu\nu} \mp
\frac{i}{2} e^{-1}\epsilon^{\mu\nu\rho\sigma}F^{A}_{\rho\sigma} \right) \ ,
\end{eqnarray}
where
\begin{equation}
\epsilon_{\mu\nu\rho\sigma}\equiv e^{-2} \epsilon^{\lambda\kappa\eta\theta}g_{\mu\lambda}g_{\nu\kappa}g_{\rho\eta}g_{\sigma\theta} \ ,
\end{equation}
the last two terms of (\ref{redlag1b}) can also be written as
\begin{eqnarray}
e^{-1}\mathcal{L}^{(4)\textrm{vec}}_{\textrm{kin}}&=&\frac{1}{2}\textrm{Im}(\mathcal{N}_{AB}F_{\mu\nu}^{A+}F^{\mu\nu B+})\nonumber\\
&\equiv&-\frac{i}{4}(\mathcal{N}_{AB}F_{\mu\nu}^{A+}F^{\mu\nu B+}-
\bar{\mathcal{N}}_{AB} F_{\mu\nu}^{A-}F^{\mu\nu B-}) .
\label{vectorkinetic}
\end{eqnarray}


\subsubsection{Symplectic reparametrization and global symmetries} \label{symsec}
The field strengths $F_{\mu\nu}^{A +}$ and their
 ``duals'',
\begin{equation}
G_{\mu\nu A}^{+}:=\frac{\delta \mathcal{L}^{(4)}}{\delta F_{\mu\nu}^{A+}}=-\frac{ie}{2}\mathcal{N}_{AB}F^{\mu\nu B + },
\end{equation}
can be combined into a symplectic vector
\begin{equation} \label{symsec2}
 \left( \begin{array}{c}
 F_{\mu\nu}^{A +} \\
 G_{\mu\nu B}^{ + }
 \end{array}  \right)
\end{equation}
so that the equations of motion that follow from (\ref{redlag1b})
are invariant under the global symplectic rotations
\begin{equation}
 \left( \begin{array}{c}
 X^{A} \\
 F_{B}
 \end{array}  \right)  \longrightarrow  \mathcal{O} \left( \begin{array}{c}
 X^{A} \\
 F_{B}
 \end{array}  \right)  , \qquad   \left( \begin{array}{c}
 F_{\mu\nu}^{A +} \\
 G_{\mu\nu B}^{ + }
 \end{array}  \right)   \longrightarrow  \mathcal{O}
   \left( \begin{array}{c}
 F_{\mu\nu}^{A +} \\
 G_{\mu\nu B}^{ + }
 \end{array}  \right)
\end{equation}
with $\mathcal{O}$ being  a symplectic matrix with respect to the
symplectic metric
\begin{equation}
\omega =  \left( \begin{array}{cc}
0 & \delta_{B}{}^{A} \\
- \delta^{C}{}_{D} & 0
\end{array}  \right)  .
\end{equation}
namely $\mathcal{O}^T \omega \mathcal{O} = \omega.$ Writing
$\mathcal{O}$ as
\begin{equation}
\mathcal{O} =   \left(  \begin{array}{cc}
A & B \\
C & D
  \end{array}      \right)  ,
\end{equation}
the period matrix $\mathcal{N}$ transforms as
\begin{equation}
\mathcal{N} \longrightarrow  (C + D \mathcal{N} ) ( A + B \mathcal{N} )^{-1}  .
\label{Ntrafo}
\end{equation}

Symplectic transformations with $B\neq 0$ correspond to non-perturbative
electromagnetic duality transformations,  whereas transformations with
$C\neq 0$ involve shifts of the theta angles in the Lagrangian.

General symplectic tranformations will take a Lagrangian
$\mathcal{L}(F,G)$ with the field strengths
satisfying the Bianchi identities $dF^{A}=0$ and
$dG_{A}=0$ to a Lagrangian
$\tilde{\mathcal{L}}(\tilde{F},\tilde{G})$ with
the new field strengths satisfying $d{\tilde{F}}^{A}=0$ and
$d{\tilde{G}}_{A}=0.$

The subgroup, $\mathcal{U}$, of $Sp(2(\tn+2),\mathbb{R})$ that
leaves the functional invariant
\[
\tilde{\mathcal{L}}(\tilde{F},\tilde{G})=\mathcal{L}(\tilde{F},\tilde{G}),
\]
is called  the duality invariance group (or ``U-duality group").
This is a symmetry group of the equations of motion, and we will
call theories related by transforations in $\mathcal{U}$ ``on-shell
equivalent". A subgroup of the duality invariance group that leaves
the off-shell Lagrangian invariant up to surface terms is called an
``electric subgroup", $G_{E}$, since it transforms electric field
strengths into electric field strengths only. Obviously, we have the
inclusions \[ G_{E}\subset \mathcal{U}\subset
Sp(2(\tn+2),\mathbb{R}).\] Pure electric-magnetic exchanges are
contained in the coset $\mathcal{U}/G_{E}$. Hodge-dualizations,
contained in $Sp(2\tn+4)/\mathcal{U}$~\cite{dWSTb}, lead to ``dual
theories" that generally have different manifest electric subgroups
$G_{E}$.

A four-dimensional MESGT that derives from five dimensions with
the prepotential (\ref{prepot}) automatically has the following (global) duality
 symmetries: \footnote{There might be
 additional  hidden symmetries
for certain  scalar manifolds, such as symmetric spaces \cite{GST1},
  or some homogeneous
 spaces \cite{dWVP3}.
However, in general, there are no additional hidden symmetries. The
number of hidden symmetry generators is maximum for symmetric target
spaces in four dimensions and is equal to the number of translation
(shift) generators. }
\begin{enumerate}
\item  The whole global symmetry group  of the  5D Lagrangian, i.e., the invariance
group $G$ of the
 cubic polynomial
$\mathcal{V}(h)=C_{\ti\tj\tk} h^{\ti} h^{\tj} h^{\tk}$, survives as
a global symmetry group of the 4D theory.
\item The shifts $z^{\ti}  \rightarrow   z^{\ti} + b^{\ti}$ with constant real parameters $b^{\ti}$ (i.e., the shifts of the Kaluza-Klein
 scalars $A^{\ti}$) become symmetries of the 4D theories if they are accompanied
 by simultaneous transformations
 \begin{equation}
 F_{\mu\nu}^{\tI}  \rightarrow    F_{\mu\nu}^{\tI} - 2W_{\mu\nu}b^{\ti}
 \end{equation}
 of the field strengths.
 \item There is an additional global scaling symmetry
 \begin{equation}
 X^{0}  \rightarrow             e^{\beta} X^{0} , \qquad \qquad
 X^{\ti}   \rightarrow   e^{-\frac{\beta}{3}}  X^{\ti}
 \end{equation}
 which leaves the prepotential (\ref{prepot}) invariant.
 \end{enumerate}
 Together    these symmetries  form the global invariance group
 \begin{equation}
 (G\times SO(1,1))\ltimes T^{(\tn +1)},
 \end{equation}
where $SO(1,1)$ describes the scaling symmetry, $T^{(\tn+1)}$ refers
to the real translations of scalars by $b^{\tI}$, and $\ltimes$
denotes a semi-direct product. The symplectic matrix $\mathcal{O}$
that implements these symmetries on the symplectic sections
(\ref{symsec1}), (\ref{symsec2})
 is block diagonal for $G\times SO(1,1)$, but involves
shifts of the theta angles for the translations $T^{(\tn+1)}$. More
precisely, an infinitesimal  transformation of $(G\times
SO(1,1))\ltimes T^{(\tn +1)}$ is represented by
\begin{equation}
\mathcal{O}= \mathbf{1} +  \left(  \begin{array}{cc}
B & 0 \\
C & -B^{T}
\end{array}  \right)
\end{equation}
with
\begin{equation}
B^{A}{}_{B} =     \left(  \begin{array}{cc}
\beta & \hspace{4mm}  0\\
b^{\tI} & \hspace{4mm}  [ M_{(r)\tJ}^{\tI} +\frac{1}{3}\beta\delta_{\tJ}^{\tI}]
      \end{array}            \right)  , \qquad
      C_{AB}   =    \left(  \begin{array}{cc}
      0 & \hspace{4mm}  0 \\
      0 & \hspace{4mm}   -2\sqrt{2} C_{\tI\tJ\tK}b^{\tK}
      \end{array}     \right)  , \label{symmat2}
\end{equation}
 where $b^{\ti}$ is now an infinitesimal shift parameter and only
terms linear in the transformation parameters are kept. Note that
for different symplectic sections, the above transformation matrices
also change their form in general. In order to gauge symmetries in
the standard way \cite{S4,dWLVP}, one works in a symplectic basis, where
the symmetries one wants     to gauge are represented by
block-diagonal symplectic matrices. However, there are
 cases in which
also off-diagonal transformations can be gauged by  certain ``non-standard''
gaugings \cite{dWLVP,dWHR}, but often  these gaugings turn out to be dual to a standard gauging.
We will come back to this point later.


\subsection{The dimensional reduction of  $\mathcal{N}=2$ YMESGTs with tensor fields}
In this subsection, we consider the dimensional reduction of a  5D
 YMESGT  with tensor fields  to 4D. Our starting point is thus the
5D Lagrangian (\ref{start}). Just as for the ungauged case, we
 decompose the f\"{u}nfbein as in eq. (\ref{fuenfbein})
 and the vector fields $\hat{A}_{\hat{\mu}}^{I}$
 as in (\ref{vector}) (remembering that we now no longer
 have 5D  vector fields  with an index $M$, as these are
 converted to  5D   tensor fields).
   The 5D tensor fields $\hB_{\hmu\hnu}^{M}$
      are decomposed into      Kaluza-Klein invariant
  4D tensor fields, $B_{\mu\nu}^{M}$, and 4D vector fields, $B_{\mu}^{M}$:
\begin{equation}
\hB_{\hmu\hnu}^{M}=              \left(   \begin{array}{c} \hB_{\mu\nu}^{M}\\
\hB_{\mu 5}^{M} \end{array}    \right)   = \left( \begin{array}{c}   B_{\mu\nu}^{M}-4W_{[\mu}B_{\nu]}^{M}\\    B_{\mu}^{M}   \end{array}  \right). \label{tensor}
 \end{equation}
      As is outlined in  Appendix A,  this results in the 4D Lagrangian
 \begin{eqnarray}
 e^{-1}\mathcal{L}^{(4)} &=&-\frac{1}{2}R -\frac{3}{4}
 {\stackrel{\circ}{a}}_{\tI\tJ}(\mathcal{D}_{\mu}\tilde{h}^{\tI})(\mathcal{D}^{\mu}\tilde{h}^{\tJ})
-\frac{1}{2}e^{-2\sigma}{\stackrel{\circ}{a}}_{IJ}(\mathcal{D}_{\mu} A^{I})(\mathcal{D}^{\mu}A^{J}) \nonumber\\
&&-e^{-2\sigma}{\stackrel{\circ}{a}}_{IM}(\mathcal{D}_{\mu}A^{I})B^{\mu M}
-\frac{1}{2}e^{-2\sigma}{\stackrel{\circ}{a}}_{MN} B_{\mu}^{M}B^{\mu N} \nonumber\\
&&+\frac{e^{-1}}{g} \epsilon^{\mu\nu\rho\sigma}\Omega_{MN}B_{\mu\nu}^{M}
(\partial_{\rho} B_{\sigma}^{N}+ g A_{\rho}^{I}\Lambda_{IP}^{N}B_{\sigma}^{P})\nonumber\\
& &  +\frac{e^{-1}}{g} \epsilon^{\mu\nu\rho\sigma}\Omega_{MN}W_{\mu\nu}B_{\rho}^{M}B_{\sigma}^{N} +\frac{e^{-1}}{2\sqrt{6}} C_{MNI}\epsilon^{\mu\nu\rho\sigma}B_{\mu\nu}^{M}
B_{\rho\sigma}^{N} A^{I}\nonumber\\
& &  -\frac{1}{4}
e^{\sigma}{\stackrel{\circ}{a}}_{MN}B_{\mu\nu}^{M}B^{N\mu\nu}
-\frac{1}{2}e^{\sigma}{\stackrel{\circ}{a}}_{IM}(\mathcal{F}_{\mu\nu}^{I}+2W_{\mu\nu}A^{I})B^{M\mu\nu}\nonumber\\
& &  -\frac{1}{4}e^{\sigma}{\stackrel{\circ}{a}}_{IJ}(\mathcal{F}_{\mu\nu}^{I}+2W_{\mu\nu}A^{I}
 )( \mathcal{F}^{J\mu\nu}+2W^{\mu\nu}A^{J})  -\frac{1}{2}e^{3\sigma}W_{\mu\nu}W^{\mu\nu} \nonumber \\
   & &   +\frac{e^{-1}}{2\sqrt{6}} C_{IJK}\epsilon^{\mu\nu\rho\sigma} \Big\{
\mathcal{F}_{\mu\nu}^{I}\mathcal{F}_{\rho\sigma}^{J}A^{K} + 2 \mathcal{F}_{\mu\nu}^{I}W_{\rho\sigma}A^{J}A^{K} +\frac{4}{3}W_{\mu\nu}W_{\rho\sigma} A^{I}A^{J}A^{K} \Big\} \nonumber\\
       & & -g^2 P,  \label{redlag2}
\end{eqnarray}
where\begin{eqnarray}
\mathcal{D}_{\mu}A^{I} & \equiv  & \partial_{\mu} A^{I} +g A_{\mu}^{J}f_{JK}^{I}A^{K}\\
\mathcal{F}_{\mu\nu}^{I}  &  \equiv  &    2\partial_{[\mu}A_{\nu]}^{I} + gf_{JK}^{I}A_{\mu}^{J}A_{\nu}^{K} \\
\mathcal{D}_{\mu}\tilde{h}^{\tI}& \equiv & \partial_{\mu} \tilde{h}^{\tI} + g A_{\mu}^{I}M_{I\tK}^{\tI}\tilde{h}^{\tK},
\end{eqnarray}
and the total scalar potential, $P$, is given by
\begin{equation}
P=e^{-\sigma}P^{(T)} +\frac{3}{4} e^{-3\sigma}{\stackrel{\circ}{a}}_{\tI\tJ} (A^{I}M_{I\tK}^{\tI} h^{\tK})
(A^{J}M_{J\tilde{L}}^{\tJ}h^{\tilde{L}}), \label{totalpot}
\end{equation}
Note that, in the first line of (\ref{redlag2}),  we have absorbed the kinetic term of sigma by defining $\tilde{h}^{\ti}$ as in (\ref{htildeI}).

This Lagrangian has several interesting features:
\begin{itemize}
\item Whereas the scalars $h^{\tI}$ are complete, the scalars $A^{M}$
one had in the ungauged theory, have disappeared from the Lagrangian.
This was to be expected, as the scalars $A^{M}$ in the ungauged theory have their origin in  the 5D   vector fields $\hat{A}_{\hmu}^{M}$, which, however,  are dualized
to  the   5D  two-form fields $\hat{B}_{\hmu\hnu}^{M}$ in the gauged version,
 and the $\hat{B}_{\hmu\hnu}^{M}$  do not give rise to  4D scalar fields.
\item The terms in the second line of (\ref{redlag2})
  suggest that the scalar $A^{M}$ has been
eaten by the vector fields $B_{\mu}^{M}$ as the result     of a Peccei-Quinn-type       gauging of the translations $A^{M}       \rightarrow  A^{M} + b^{M}$ (cf. sec.  \ref{symsec}). In the standard symplectic basis,
however, the shifts of the $A^{\tI}$ are not blockdiagonal symplectic transformations (see eq. (\ref{symmat2})). The conventional gauging
of isometries of the scalar manifold described in \cite{S4,dWLVP} requires a blockdiagonal embedding
of the isometries in the corresponding symplectic duality matrices.
The theory at hand can therefore be interpreted in either of two ways:
either as a \emph{non}-standard gauging in the ``conventional''
symplectic basis, or as a standard gauging in a rotated symplectic section.
We will map the above theory to such a standard gauging below.
\item ``Regurgitating''  scalar fields $A^{M}$ from the $B^{M}_{\mu}$, or, more precisely, making the replacement
\begin{equation}
B_{\mu}^{M} \rightarrow gB_{\mu}^{M} +\mathcal{D}_{\mu}A^{M}
\end{equation}
together with the  shift
\begin{equation}
B_{\mu\nu}^{M} \rightarrow gB_{\mu\nu}^{M} + F_{\mu\nu}^{M}+2W_{\mu\nu}A^{M},
\end{equation}
and switching off $g$, leads back to the ungauged  theory (\ref{redlag}).
\item After having eaten the scalar fields $A^{M}$, the vector fields
$B_{\mu}^{M}$ acquire a mass term (the last term in the second line
of (\ref{redlag2})). However, there is no explicit \emph{kinetic} term for
the $B_{\mu}^{M}$. Instead, there  are the two-form fields
$B_{\mu\nu}^{M}$, which also have a mass term (in  line 5 of
(\ref{redlag2})), but also no second order kinetic term. The
two-forms have  a  one derivative interaction with the  vectors
$B_{\mu}^{M}$ in the third line of (\ref{redlag2}). Such a term
normally allows the elimination of  the tensor fields in favor of
the vector fields or vice versa, so that one either obtains massive
vector fields with a standard second order kinetic term or massive
tensor fields with a  standard second order kinetic term. This is
possible because massive vectors are dual to massive tensors in 4D.
As we will show below, it is indeed possible to eliminate the
tensors $B_{\mu\nu}^{M}$ in favor of the vectors $B_{\mu}^{M}$.
However, the converse seems to be difficult, if not impossible  to
achieve locally, due to the term proportional to
$\epsilon^{\mu\nu\rho\sigma}W_{\mu\nu}B_{\rho}^{M}B_{\sigma}^{N}$ in
the fourth line in (\ref{redlag2}).
\item The tensors $B_{\mu\nu}^{M}$ and the vectors $B_{\mu}^{M}$ are
charged under the 5D gauge group, $K$, which also descends to a local gauge symmetry in 4D.
This is in contrast to the massive tensor fields that have been recently considered in the
literature \cite{T42,T41,T43}. The tensor fields in those papers arise from dualizations of scalars
 of the quaternionic manifold instead of the special K\"{a}hler manifold and also don't carry
 any charge with respect to a non-trivial local gauge group.  The Lagrangian (\ref{redlag2}) does, however,
have some resemblance with   the Freedman-Townsend model \cite{FT}    (see also \cite{new}).
\item The terms in the sixth and seventh line of (\ref{redlag2}) can be written as
\begin{equation}
\frac{1}{2}\textrm{Im}\Big[\mathcal{N}_{00}F_{\mu\nu}^{0+}F^{\mu\nu 0 + }
+2\mathcal{N}_{0I}F_{\mu\nu}^{0 +} \mathcal{F}^{\mu\nu I +} +\mathcal{N}_{IJ} \mathcal{F}_{\mu\nu}^{ I + }
 \mathcal{F}^{\mu \nu  I + } \Big]\Big|_{A^{M}=0}.
\end{equation}
\end{itemize}



\section{Eliminating the tensor fields}
The action (\ref{redlag2}) contains the terms
\begin{equation}
-\frac{1}{4}
e^{\sigma}{\stackrel{\circ}{a}}_{MN}B_{\mu\nu}^{M}B^{N\mu\nu}
-\frac{1}{2}e^{-2\sigma}{\stackrel{\circ}{a}}_{MN} B_{\mu}^{M}B^{\mu N}
+\frac{e^{-1}}{g} \epsilon^{\mu\nu\rho\sigma}\Omega_{MN}B_{\mu\nu}^{M}
\partial_{\rho} B_{\sigma}^{N}
\end{equation}
If these were the only terms involving $B_{\mu\nu}^{M}$ and $B_{\mu}^{N}$,
one could simply,  as mentioned in the previous section,  integrate out
$B_{\mu\nu}^{M}$ in favor of $B_{\mu}^{N}$, which, schematically,
   would result   in a relation of the form
\begin{equation}
B^{\mu\nu M} = \mathcal{T}^{M}{}_{N} e^{-1}  \epsilon^{\mu\nu\rho\sigma}\partial_{\rho}B_{\sigma}^{N}
\label{B2B1}
\end{equation}
with some matrix $\mathcal{T}^{M}{}_{N}$ and leads
 to  a standard second
order action for  massive vector fields $B_{\mu}^{N}$,
\begin{equation}
-\mathcal{K}_{MN}(\partial_{[\mu}B_{\nu]}^{M})(\partial^{[\mu}B^{\nu]N})   - \mathcal{M}_{MN}B_{\mu}^{M}B^{\mu N},
\end{equation}
with a kinetic  and a mass matrix $\mathcal{K}_{MN}$ and $\mathcal{M}_{MN}$,
respectively.

Alternatively,  one could also choose to integrate out the
vector fields in favor of the tensor fields, which then would lead to
a relation of the form
\begin{equation}
B^{\mu M} = \tilde{\mathcal{T}}^{M}{}_{N} e^{-1}  \epsilon^{\mu\nu\rho\sigma}\partial_{\nu}B_{\rho\sigma}^{N}\label{B1B2}
\end{equation}
and a standard second order action for massive tensor fields,
\begin{equation}
-\tilde{\mathcal{K}}_{MN}(\partial_{[\mu}B_{\nu\rho]}^{M})( \partial^{[\mu}B^{\nu\rho]N})- \tilde{\mathcal{M}}_{MN}B_{\mu\nu}^{M}B^{\mu\nu N}.
\end{equation}
However, this is not quite what happens, as
in the actual Lagrangian (\ref{redlag2}),  there are also
other quadratic
terms of the form
\begin{equation}
\frac{e^{-1}}{g} \Omega_{MN}\epsilon^{\mu\nu\rho\sigma} W_{\mu\nu} B_{\rho}^{M}B_{\sigma}^{N}
\end{equation}
  and
  \begin{equation}
\frac{e^{-1}}{2\sqrt{6}}  C_{MNI}\epsilon^{\mu\nu\rho\sigma}B_{\mu\nu}^{M}B_{\rho\sigma}^{N}A^I.
\label{BBA}
  \end{equation}
The first of these two terms would contribute a term proportional
to
\begin{equation}
e^{-1}  \epsilon^{\mu\nu\rho\sigma}W_{\nu\rho}B_{\sigma}^{M}
\end{equation}
to  the left hand side of eq. (\ref{B1B2}).
This additional term   seems  to make it
impossible to (locally)  eliminate
the vector fields $B_{\mu}^{N}$ in favor of the tensor fields
$B_{\mu\nu}^{M}$, as  the field strength $W_{\mu\nu}$
would somehow have to be ``inverted'' to solve the  equation  for
$B_{\mu}^{N}$.
The second term, (\ref{BBA}), on the other hand, would yield a contribution
involving
\begin{equation}
e^{-1}  C_{MNI}A^I\epsilon^{\mu\nu\rho\sigma}B_{\rho\sigma}^{N}
\end{equation}
to          the  left hand side of (\ref{B2B1}). This involves only scalar fields
in front of $B_{\mu\nu}^{M}$, which, in principle, can be inverted
so as to solve the modified eq. (\ref{B2B1}) for $B_{\mu\nu}^{M}$. Due to the epsilon tensor, however,
one has to switch to the selfdual and anti-selfdual components
of all two-forms. In addition, there are also further terms linear in the $B_{\mu\nu}^{M}$ in eq.
(\ref{redlag2}), which we have neglected in the above schematic discussion.
Let us therefore become more specific now and carry out the elimination of the tensor fields in detail.
 To this end, we write the $B_{\mu\nu}^{M}$-dependent terms in (\ref{redlag2}) as follows
\begin{equation}
e^{-1}  \mathcal{L}_{B_{\mu\nu}^{M}}^{(4)}=\frac{1}{2}\textrm{Im}
\Big[\mathcal{N}_{MN} B_{\mu\nu}^{M + } B^{\mu\nu N + } \Big]\Big|_{A^{M}=0}
+2 \textrm{Re} \Big[ J_{M}^{+\mu\nu } B_{\mu\nu}^{ M + } \Big] , \label{LB}
\end{equation}
where we have introduced
\begin{equation}
J^{\mu\nu}_{M} := -\frac{1}{2}  e^{\sigma} {\stackrel{\circ}{a}}_{IM}
(\mathcal{F}^{\mu\nu I} + 2 W^{\mu\nu}A^{I}) +\frac{e^{-1}}{g}\epsilon^{\mu\nu\rho\sigma}\Omega_{MN}\mathcal{D}_{\rho}B_{\sigma}^{N}
\end{equation}
and used
\begin{equation}
\mathcal{N}_{MN}\big|_{A^{M}=0} = -\frac{4}{\sqrt{6}} C_{MNI} A^{I} -i e^{\sigma} {\stackrel{\circ}{a}}_{MN}.
\end{equation}
Varying        (\ref{LB})     with    respect to $B_{\mu\nu}^{M}$, one obtains
\begin{equation}
J_{M}^{\mu\nu + }=\frac{i}{2}\mathcal{N}_{MN}\big|_{A^{M}=0} B^{\mu\nu  N + },
\end{equation}
which can be used to express $B_{\mu\nu}^{M + } $ in terms of $J_{\mu\nu M}^{ + } $ in (\ref{LB}) with the result
\begin{equation}
e^{-1}  \mathcal{L}_{B_{\mu\nu}^{M}}^{(4)} = 2 \textrm{Im} \Big[   \mathcal{N}^{ M N } J_{\mu\nu M}^{ + } J_{N}^{ \mu\nu + } \Big]\Big|_{A^{M}=0}.
\end{equation}
Here, $\mathcal{N}^{ M N }$ denotes the inverse of
$\mathcal{N}_{ M N }$,
\begin{equation}
\mathcal{N}_{ M N } \mathcal{N}^{  N  P    } = \delta_{M}^{  P  }.
\end{equation}
The Lagrangian (\ref{redlag2}) now takes on a more concise form:
\begin{eqnarray}
e^{-1}\mathcal{L}^{(4)} &=&-\frac{1}{2}R -\frac{3}{4}
{\stackrel{\circ}{a}}_{\tI\tJ}(\mathcal{D}_{\mu}\tilde{h}^{\tI})(\mathcal{D}^{\mu}\tilde{h}^{\tJ})
-\frac{1}{2}e^{-2\sigma}{\stackrel{\circ}{a}}_{IJ}(\mathcal{D}_{\mu} A^{I})(\mathcal{D}^{\mu}A^{J}) \nonumber\\
&&-e^{-2\sigma}{\stackrel{\circ}{a}}_{IM}(\mathcal{D}_{\mu}A^{I})B^{\mu M}
-\frac{1}{2}e^{-2\sigma}{\stackrel{\circ}{a}}_{MN} B_{\mu}^{M}B^{\mu N} \nonumber\\
& &+\frac{1}{2}\textrm{Im}\Big[\mathcal{N}_{00}F_{\mu\nu}^{0+}F^{\mu\nu 0 + }
+2\mathcal{N}_{0I}F_{\mu\nu}^{0 +} \mathcal{F}^{\mu\nu I +} +\mathcal{N}_{IJ} \mathcal{F}_{\mu\nu}^{ I + } \mathcal{F}^{\mu \nu  I + } \Big]\Big|_{A^{M}=0} \nonumber\\
& &+ 2 \textrm{Im} \Big[   \mathcal{N}^{ M N } J_{\mu\nu M}^{ + } J_{N}^{ \mu\nu + } \Big]\Big|_{A^{M}=0}  +\frac{e^{-1}}{g} \epsilon^{\mu\nu\rho\sigma}\Omega_{MN}W_{\mu\nu}B_{\rho}^{M}B_{\sigma}^{N}  \nonumber \\
 & & -g^2 P.  \label{redlag3}
\end{eqnarray}



\section{The equivalence to a standard gauging}
In this section, we show that the above action (\ref{redlag3})
is dual to a standard gauged supergravity theory of the type described in \cite{S4,dWLVP}.
We already identified the translations $A^M  \rightarrow   b^M$
and the 5D gauge group $K$ generated by the matrices $M_{(I)\tJ}^{\tI}$
of eq. (\ref{MfB}) as  part of
the 4D gauge group. We also saw,      in     (\ref{symmat2}), however, that, in the ungauged theory,
 the translations
$A^{\tI} \rightarrow b^{\tI} $ are not represented by block diagonal
symplectic matrices if one works in the ``natural'' symplectic basis
\begin{equation}
 \left( \begin{array}{c}
 X^A \\
 F_{B}
 \end{array} \right) = \left( \begin{array}{c}
 X^{A} \\
 \partial_{B} F  \end{array}  \right) \label{vec1}
 \end{equation}
with $
X^{0} = 1$ and $X^{\ti} = z^{\ti} $
and
\begin{equation}
\left( \begin{array}{c}
F_{\mu\nu}^{A} \\
G_{\mu\nu B}
\end{array}   \right)  ,
\end{equation}
with $F_{\mu\nu}^{0}=-2\sqrt{3} W_{\mu\nu}$,
which one directly gets from  the dimensional reduction from 5D.
In order to gauge the translations associated with $b^{M}$
in the standard way,  one therefore has
to go  to a different symplectic basis in which both the translations
by $b^{M}$ and the $K$  transformations are represented by block diagonal
symplectic matrices.
To see how this works, we split the $z^{\ti}$ into $(z^{I},z^{M})$
and take into account that $C_{MNP}=C_{IJM}=0$. The symplectic vector
(\ref{vec1}) then becomes
\begin{equation}
\left( \begin{array}{c}
X^{0}\\
X^{I}\\
X^{M}\\
F_{0}\\
F_{I}\\
F_{M}
\end{array} \right) = \left( \begin{array}{c}
1\\z^{I}\\z^{M}\\
\sqrt{2}/3 [C_{IJK}z^{I}z^{J}z^{K} +3 C_{IMN}z^{I}z^{M}z^{N}]\\
-\sqrt{2} [C_{IJK}z^{J}z^{K}+C_{IMN}z^{M}z^{N}]\\
-2\sqrt{2} C_{MNI} z^{N}z^{I}
\end{array} \right) \label{normbasis}
\end{equation}
Under an  infinitesimal translation $z^{M} \rightarrow z^{M} + b^{M}$, this transforms
as
\begin{equation}
\left( \begin{array}{c}
X^{0}\\
X^{I}\\
X^{M}\\
F_{0}\\
F_{I}\\
F_{M}
\end{array} \right) \rightarrow
\left( \begin{array}{c}
X^{0}\\
X^{I}\\
X^{M}\\
F_{0}\\
F_{I}\\
F_{M}
\end{array} \right) + \left( \begin{array}{c}
0\\
0\\
b^M X^{0} \\
 -b^M F_{M}  \\
-       2    \sqrt{2}b^{M}C_{MNI}X^{N} \\
-       2    \sqrt{2}b^{N}C_{MIN}X^{I}
\end{array} \right) ,
\end{equation}
where we have, somewhat redundantly, inserted $X^{0}=1$ in the third line
and kept only terms linear in  the infinitesimal parameters $b^{M}$.

From this expression, it becomes clear that $(X^{0}, F_{I}, X^{M})$
transform among themselves, as do $(F_{0}, X^{I},  F_{M})$. Thus, a symplectic
duality rotation that exchanges $X^{0}$ with $F_{0}$ and $X^{M}$ with $F_{M}$,
       could  make the translations $z^{M} \rightarrow z^{M} + b^{M}$ blockdiagonal.
At the same time, we want this symplectic duality rotation to preserve the
       block diagonality       of the $K$ transformations (\ref{MfB}).
In our original basis (\ref{normbasis}), a  combined infinitesimal
translation  $z^{M} \rightarrow z^{M} + b^{M}$ and infinitesimal   $K$
transformation with parameter $\alpha^I$ is generated by the
symplectic matrix
       \begin{equation}
 \mathcal{O}  = \mathbf{1} +  \left(  \begin{array}{cc}
 B & 0 \\
 C & -B^{T}
 \end{array}  \right)  ,
 \end{equation}
  with
\begin{equation}
B =  \left( \begin{array}{ccc}
0 & 0 & 0 \\
0 & \alpha^I f_{IJ}^{K} & 0\\
b^{M} & 0 & \alpha^I \Lambda_{IN}^{M}
 \end{array}   \right)   , \qquad C =
   \left(    \begin{array}{ccc}
   0&0&0\\
   0 & 0 & B_{IM}\\
   0 & B_{MI} & 0 \end{array}
      \right)   ,
\end{equation}
where
\begin{equation}
B_{IM} := -2 \sqrt{2} C_{IMN}b^{N}.
\end{equation}
In order to get   this block diagonal,  we     switch to a new symplectic basis
\begin{equation}
  \left(  \begin{array}{c}
  X^{A} \\
  F_{B}
  \end{array}   \right)   \rightarrow    \left(  \begin{array}{c}
  \check{X}^{A} \\
  \check{F}_{B}
  \end{array}   \right)   \equiv    \mathcal{S}   \left(  \begin{array}{c}
  X^{A} \\
  F_{B}
  \end{array}   \right) , \qquad  \left(  \begin{array}{c}
  F_{\mu\nu}^{A} \\
  G_{\mu\nu B}
  \end{array}   \right)   \rightarrow    \left(  \begin{array}{c}
  \check{F}_{\mu\nu}^{A} \\
  \check{G}_{\mu\nu B}
  \end{array}   \right)   \equiv  \mathcal{S}   \left(  \begin{array}{c}
  F_{\mu\nu}^{A} \\
  G_{\mu\nu B}
  \end{array}   \right)
\end{equation}
where
\begin{equation}
\mathcal{S} =  \left(  \begin{array}{cccccc}
0 & 0 & 0 & 1 & 0 & 0 \\
0 & \delta^{J}{}_{I} & 0 &0& 0& 0\\
0 & 0 & 0 & 0 & 0 & D^{MN}\\
-1 & 0 & 0 & 0 & 0 & 0 \\
0 & 0& 0& 0& \delta_{I}{}^{J}& 0 \\
0 & 0 & D_{MN}  &0&0&0
\end{array}
\right) ,\label{Sdef}
\end{equation}
and
\begin{equation}
D_{MN}:=-2\sqrt{3} \Omega_{MN}, \qquad D_{MN}D^{NP}=\delta_{M}^{P} . \label{Ddef}
\end{equation}
It is easy   to verify that the rotation matrix $\mathcal{S}$ is itself symplectic
and that
\begin{equation}
\check{\mathcal{O}}\equiv \mathcal{S} \mathcal{O} \mathcal{S}^{-1} =
\mathbf{1}  +   \left(   \begin{array}{cc}
\check{B} & 0 \\
0 &  - \check{B}^{T}
\end{array}      \right) \label{blockdiagonal}
\end{equation}
with
\begin{equation}
\check{B}     = \left(   \begin{array}{ccc}
0 & 0 &  2\sqrt{3} b^{M} \Omega_{MN} \\
0 & \alpha^I f_{IJ}^{K} & 0 \\
0 & + \Lambda_{IM}^{N} b^{M} & \alpha^I \Lambda_{IM}^{N}
\end{array}   \right)  \label{block} .
\end{equation}

Here, we have used (\ref{OL1}), (\ref{OL2}) and (\ref{Ddef}). Hence,
in the new basis $(\check{X}^{A}, \check{F}_{B})$,
$(\check{F}_{\mu\nu}^{A}, \check{G}_{\mu\nu B} )$, the group $K
\ltimes \mathbb{R}^{n_{T}} $ is represented by block diagonal
symplectic matrices $\check{O}$.
 But this is not all; setting
\begin{equation}
\check{B}^{C}{}_{B} = \alpha^{A}f_{AB}^{C} ,
\end{equation}
one reads off
\begin{equation}
f_{IJ}^{K}, \qquad f_{IM}^{N} = \Lambda_{IM}^{N} = - f_{MI}^{N},
\qquad f_{MN}^{0} = -2 \sqrt{3} \Omega_{MN} ,  \label{Of}
\end{equation}
 as the non-vanishing components as well as
 $\alpha^M= -b^M$.
It is easy to see that the non-vanishing $f_{AB}^{C}$ of eq. (\ref{Of})
define  the   Lie algebra of  a central extension of
the   Lie algebra  of
 $K\ltimes   \mathbb{R}^{n_{T}}$, with the central charge   corresponding to the index $0$.\footnote{Note that there is a subtlety here concerning the central charge.   As one easily verifies, two translations represented by matrices of the form  (\ref{blockdiagonal}) and (\ref{block}) with $\alpha^{I}=0$ and two parameters $b^{M}$ and $b^{M'}$,
always commute, even though $f_{MN}^{0} \neq 0$. However that is a generic property     of finite-dimensional representations of centrally extended Lie algebras such as the Heisenberg algebra.}    We shall denote the corresponding group  of this   centrally extended
Lie algebra as $K \ltimes \mathcal{H}^{n_T+1}$,   where
$\mathcal{H}^{n_T+1}$ denotes the Heisenberg group generated by the
translations and the central charge.

As the structure constants define the adjoint representation,
this centrally extended group can therefore be gauged in the
standard way    if one uses the new symplectic basis $(\check{X}^{A},\check{F}_{B})$.   As we will show now, the resulting Lagrangian     of this  $K \ltimes \mathcal{H}^{n_T+1}$  gauged theory  is dual to the Lagrangian (\ref{redlag3}) of the
previous section,  which we got from the dimensional reduction of a 5D theory with tensor fields.   In order to show this, we will start from the    4D   ungauged
  theory  in the
new symplectic basis $(\check{X}^{A},\check{F}_{B})$, $(\check{F}_{\mu\nu}^{A},\check{G}_{\mu\nu B})$  and assume the subsequent  gauging of the group $K\ltimes \mathcal{H}^{n_{T}+1}$ using the standard formulae \cite{S4,dWLVP} evaluated in that new basis. As this gauging is
fairly standard, we  can    skip the details and  immediately  write down  the   resulting Lagrangian.
This standard Lagrangian with gauge group $K\ltimes \mathcal{H}^{n_{T}+1}$  will then be  subjected to a few field redefinitions
and dualizations until it precisely coincides with the Lagrangian (\ref{redlag3}) from the dimensional reduction of a 5D theory with tensor fields.

We will first carry out this program for  the scalar sector and after that for the kinetic terms of the vector fields.

\subsection{The scalar sector}
The K\"{a}hler potential $K(z,\bar{z})$ of eq. (\ref{symK})  is a
symplectic invariant. Thus, the metric $g_{\ti\bar{\tj}}$ stays the
same as in the old symplectic basis. The gauging of  $K \ltimes
\mathcal{H}^{n_T+1}$ , however, leads to two modifications in the
scalar sector. First, the kinetic term of the scalars becomes
covariant with respect to the gauge group:
\begin{equation}
-g_{\ti\bar{\tJ}} (\partial_{\mu} z^{\tI}) (\partial^{\mu} \bar{z}^{\tJ})
\rightarrow -g_{\ti\bar{\tJ}} (\mathcal{D}_{\mu} z^{\tI}) (\mathcal{D}^{\mu} \bar{z}^{\tJ})
\end{equation}
with
\begin{equation}
\mathcal{D}_{\mu} z^{\tI} = \partial_{\mu} z^{\tI} +g \check{A}_{\mu}^{A}K_{A}^{\tI} .
 \end{equation}
Here, $K_{A}^{\tI}(z)$ are the holomorphic Killing vectors that generate the gauge group on the scalar manifold $\mathcal{M}^{(4)}$. They  can be expressed
 in  terms of derivatives of Killing prepotentials, $P_{A}$,
\begin{equation}
K_{A}^{\tI} = i g^{\tI\bar{\tJ}} \partial_{\bar{\tJ}} P_{A} ,
\end{equation}
where \cite{S4,dWLVP}
\begin{equation}
P_{A}=e^{K} ( \check{F}_{B} f_{AC}^{B} \bar{\check{X}}^{C} + \bar{\check{F}}_{B}f_{AC}^{B}\check{X}^{C}) .
\end{equation}
Using this, one obtains
\begin{eqnarray}
P_{0} & = & 0  \nonumber \\
P_{I} & = & -\sqrt{2}  e^{K} \Big(C_{\ti\tj\tk} z^{\tj} z^{\tk} M_{(I) \tilde{L}}^{\tI}\bar{z}^{\tilde{L}}   \Big)  + \textrm{c.c.}\nonumber \\
 P_{M} & = & -2\sqrt{2} e^{K} C_{IMP} \Big( z^{P} \bar{z}^{I}
 - \bar{z}^{P} \bar{z}^{I} \Big) + \textrm{c.c.}
\end{eqnarray}
and then
\begin{eqnarray}
K_{0}^{\tJ} & = & 0 \nonumber \\
K_{I}^{\tJ} & = & M_{(I)\tilde{L}}^{\tJ} z^{\tilde{L}}   \nonumber \\
K_{M}^{\tI} & = & -  \delta^{\tI}_{M}.  \label{VE}
\end{eqnarray}
Upon the identification
\begin{eqnarray}
A_{\mu}^{I} & = & \check{A}_{\mu}^{I} \nonumber \\
B_{\mu}^{M}&=&-g\sqrt{3} \check{A}_{\mu}^{M}, \label{BA}
\end{eqnarray}
the kinetic term of the scalars then becomes
\begin{equation}
-g_{\tI \bar{\tJ}} (\mathcal{D}_{\mu} z^{\tI}) (\mathcal{D}^{\mu} \bar{z}^{\tJ})
= -\frac{3}{4}
{\stackrel{\circ}{a}}_{\tI\tJ}(\mathcal{D}_{\mu}\tilde{h}^{\tI})(\mathcal{D}^{\mu}\tilde{h}^{\tJ})
-\frac{1}{2}e^{-2\sigma}{\stackrel{\circ}{a}}_{\tI\tJ}(\mathcal{D}^{\prime}_{\mu} A^{\tI})(\mathcal{D}^{\prime \mu}A^{\tJ})
\end{equation}
with
\begin{eqnarray}
\mathcal{D}_{\mu}\tilde{h}^{\tI} & =&  \partial_{\mu} \tilde{h}^{\tI} + g A_{\mu}^{I}M_{(I)\tK}^{\tI}\tilde{h}^{\tK} \nonumber\\
\mathcal{D}^{\prime}_{\mu}A^{\tI} & = &  \partial_{\mu} A^{\tI} + g A_{\mu}^{I}M_{(I)\tK}^{\tI}A^{\tK} + B_{\mu}^{M} \delta_{M}^{\tI} .
\end{eqnarray}
The vector fields $B_{\mu}^{M}$ can now absorb the scalars $A^{M}$, as anticipated,
and, after also adding the gravitational term,  we have reproduced
 the first two  lines of (\ref{redlag3}) \footnote{ We should perhaps emphasize that here we are discussing the gauging
 of the real translational isometries (of $Re(z^M)$). The resulting massive BPS vector supermultiplets have scalars
 given by $Im(z^M)$. This is to
 be contrasted with the dimensional reduction of 5D YMESGTs with
 noncompact gauge groups, in which the 4D vector fields associated with
 noncompact symmetries belong to massive BPS supermultiplets whose scalar
 fields are $Re(z^M)$. This is best seen by the fact that in five dimensions the
non-compact gauge fields become massive by eating the scalars which
in four dimensions correspond to the imaginary part of $z^M$.
   }

The gauging also induces a  second contribution to the scalar sector, namely a  scalar potential.
 Using the standard expressions,
 this scalar potential should be
 \begin{equation}
 V=e^{K} (\check{X}^{A}\bar{K}_{A}^{\tI}) g_{\bar{\ti}\tJ}(\bar{\check{X}}^{B}
 K_{B}^{\tJ}).
 \end{equation}
 Using (\ref{VE}) and expressing the $\check{X}^{\tI}$ in terms of the
 $z^{\tI}$, one finds that $V=P$, where $P$ is the potential (\ref{totalpot})
 of the dimensionally reduced Lagrangian (\ref{redlag3}).
 Thus, the two scalar sectors completely agree.
 It remains to verify the  agreement for the kinetic terms of the vector fields.

   \subsection{The kinetic terms of the vector fields}
We shall now compare kinetic terms of the vector fields of
(\ref{redlag3}) with those of the  $K\ltimes \mathcal{H}^{n_T+1}$ gauged
theory.  By kinetic terms of the vector fields, we mean the terms in
the third and fourth line of (\ref{redlag3}), which, using
(\ref{WF0}), (\ref{BA}) and (\ref{Of}), can be rewritten as
\begin{eqnarray}
e^{-1}  \mathcal{L}^{(4) \textrm{kin}}_{\textrm{vec}} &=&     \frac{1}{2}\textrm{Im}\Big[\mathcal{N}_{00}F_{\mu\nu}^{0+}F^{\mu\nu 0 + }
+2\mathcal{N}_{0I}F_{\mu\nu}^{0 +} \check{\mathcal{F}}^{\mu\nu I +} +\mathcal{N}_{IJ} \check{\mathcal{F}}_{\mu\nu}^{ I + } \check{\mathcal{F}}^{\mu \nu  I + } \Big]\Big|_{A^M=0} \nonumber\\
& &+ 2 \textrm{Im} \Big[   \mathcal{N}^{ M N } J_{\mu\nu M}^{ + } J_{N}^{ \mu\nu + } \Big]\Big|_{A^{M}=0}  - \textrm{Im} \Big[ F_{\mu\nu}^{0+} Z^{\mu\nu +}  \Big], \label{kin2}
\end{eqnarray}
with
\begin{eqnarray}
Z_{\mu\nu}  &:= &g f_{MN}^{0} \check{A}_{\mu}^{M} \check{A}_{\nu}^{N} = -\frac{2}{\sqrt{3}g}\Omega_{MN}B_{\mu}^{M}B_{\nu}^{N}\\
J^{\mu\nu}_{ M}&\equiv& -\frac{1}{2} e^{\sigma} {\stackrel{\circ}{a}}_{IM} \Big(
\check{\mathcal{F}}^{\mu\nu I} -\frac{1}{\sqrt{3}}   F^{\mu\nu 0}A^{I} \Big)
-\sqrt{3} e^{-1}  \epsilon^{\mu\nu\rho
\sigma}\Omega_{MN}\mathcal{D}_{\rho}\check{A}_{\sigma}^{N}.
\end{eqnarray}
Using  (cf. eq. (\ref{NIJ}))
\begin{eqnarray}
\mathcal{N}_{IM} \big|_{A^M=0} & = &   -i e^{\sigma} {\stackrel{\circ}{a}}_{IM}\\
\mathcal{N}_{0M} \big|_{A^{M}=0} & = &   \frac{i}{\sqrt{3}} e^{\sigma} {\stackrel{\circ}{a}}_{IM}A^{I}
\end{eqnarray}
as well as
\begin{equation}
\mathcal{D}_{[\mu} \check{A}_{\nu ]}^{M} =  \frac{1}{2}\check{\mathcal{F}}_{\mu\nu}^{M},
\end{equation}
and  the shorthand notation  (cf. eq. (\ref{Ddef})),
\begin{equation}
D_{MN}\equiv -2\sqrt{3} \Omega_{MN},
\end{equation}
$J^{\mu\nu }_{M}$ can be rewritten as
\begin{equation}
J^{\mu\nu}_{ M}= -\frac{i}{2} \Big( \mathcal{N}_{IM} \check{\mathcal{F}}^{\mu\nu I}
 + \mathcal{N}_{I0} F^{\mu\nu 0} \Big) \Big|_{A^{M}=0}
+\frac{e^{-1}}{4}\epsilon^{\mu\nu\rho
\sigma}\Omega_{MN}\check{\mathcal{F}}_{\rho\sigma}^{N}.
\end{equation}
Inserting this in (\ref{kin2}) and regrouping some terms, one obtains
\begin{eqnarray}
e^{-1}  \mathcal{L}^{(4) \textrm{kin}}_{\textrm{vec}} &=&
\frac{1}{2} \textrm{Im} \Big[  \Big( \mathcal{N}_{00}- \mathcal{N}_{0M}\mathcal{N}^{ MN} \mathcal{N}_{N0}
\Big) F_{\mu\nu}^{0  + } F^{\mu\nu 0     +    } +2 \Big( \mathcal{N}_{0I}-
\mathcal{N}_{IM}\mathcal{N}^{ MN} \mathcal{N}_{N0}
 \Big) \check{\mathcal{F}}_{\mu\nu}^{I +} F^{\mu\nu 0 +  } \nonumber\\
 && \qquad + \Big( \mathcal{N}_{IJ} -  \mathcal{N}_{IM}\mathcal{N}^{ MN} \mathcal{N}_{NJ} \Big) \check{\mathcal{F}}_{\mu\nu}^{I + } \check{\mathcal{F}}^{\mu\nu J + }               -2 \Big( D_{PM} \mathcal{N}^{ MN}
\mathcal{N}_{N0} \Big) \check{\mathcal{F}}_{\mu\nu}^{P  +  }  F_{\mu\nu}^{
0   +   }   \nonumber \\
  && \qquad +2      \Big(  \mathcal{N}_{IM}\mathcal{N}^{ MN}D_{NP} \Big)    \check{\mathcal{F}}_{\mu\nu}^{I  +  }  \check{\mathcal{F}}^{\mu\nu  P  +  }     +  \Big( D_{PM} \mathcal{N}^{ MN} D_{NQ} \Big) \check{\mathcal{F}}_{\mu\nu}^{P  +  }  \check{\mathcal{F}}^{
      \mu \nu   Q   +    }         \nonumber \\
&&  \qquad -2 F_{\mu\nu}^{0   +  }  Z^{\mu\nu       +  } \Big]\Big|_{A^M=0} \label{kin3}
\end{eqnarray}

Eq. (\ref{kin3}) is now our final form of the dimensionally reduced theory with tensor fields. We will now show that it is  ``dual" (modulo some field redefinitions) to a standard 4D gauged supergravity theory with the gauge group $K \ltimes \mathcal{H}^{n_T +1}$. Gauging this group in the standard way requires  working in
  the symplectic basis  $(\check{X}^{A},\check{F}_{B})$ and $(\check{F}_{\mu\nu}^{A},\check{G}_{\mu\nu B})$, as we discussed at length   at the beginning of Section 6. As we have seen in Section 6.1, the scalars $A^{M}$ can be ``eaten'' by the  vector fields $\check{A}_{\mu}^{M}$ that gauge the translations of $\mathcal{H}^{n_{T}+1}$.   Assuming these scalars to be gauged away from now on, the kinetic term of the $K\ltimes \mathcal{H}^{n_{T}+1}$ gauged theory is given by
 \begin{eqnarray}
e^{-1}  \check{\mathcal{L}}^{(4)\textrm{vec}}_{\textrm{kin}} & = &
 \frac{1}{2}\textrm{Im} \Big[ \check{N}_{AB} \check{\mathcal{F}}_{\mu\nu}^{  A  +      }  \check{\mathcal{F}}^{\mu\nu B  +  }    \Big]\Big|_{A^{M}=0} \nonumber\\
 & = & \frac{1}{2} \textrm{Im} \Big[   \check{\mathcal{N}}_{00} \check{\mathcal{F}}_{\mu\nu}^{0  + } \check{\mathcal{F}}^{\mu\nu 0     +    } +2  \check{\mathcal{N}}_{0I} \check{\mathcal{F}}_{\mu\nu}^{I +} \check{\mathcal{F}}^{\mu\nu 0 +  } \nonumber\\
 && \qquad +  \check{\mathcal{N}}_{IJ}  \check{\mathcal{F}}_{\mu\nu}^{I + } \check{\mathcal{F}}^{\mu\nu J + }  +2  \check{\mathcal{N}}_{M0}  \check{\mathcal{F}}_{\mu\nu}^{M  +  }  \check{\mathcal{F}}_{\mu\nu}^{
0   +   } \nonumber \\
       && \qquad +2        \check{\mathcal{N}}_{IM}    \check{\mathcal{F}}_{\mu\nu}^{I  +  }  \check{\mathcal{F}}^{\mu\nu  M  +   }
       +   \check{\mathcal{N}}_{ MN} \check{\mathcal{F}}_{\mu\nu}^{M  +  }  \check{\mathcal{F}}^{
      \mu \nu   N   +    }       \Big]\Big|_{A^M=0} , \label{kin4}
 \end{eqnarray}
 where $\check{\mathcal{N}}_{AB}$ is the period matrix in the basis
 $(\check{X}^{A}, \check{F}_{B} ) $     (to be worked out below),
and
\begin{equation}
  \check{\mathcal{F}}_{\mu\nu}^{C} = 2\partial_{[\mu}\check{A}_{\nu]}^{C} +gf^{C}_{AB}\check{A}_{\mu}^{A}\check{A}^{B}_{\nu}.
  \end{equation}
with the structure constants of eqs. (\ref{Of}).
Note that, due  to $f_{0A}^{B} =0  $, the vector field $\check{A}_{\mu}^{0}$ only appears
via its curl in $\check{\mathcal{F}}_{\mu\nu}^{0}$:
\begin{equation}
  \check{\mathcal{F}}_{\mu\nu}^{0} = 2\partial_{[\mu}\check{A}_{\nu]}^{0} +gf^{0}_{MN}\check{A}_{\mu}^{M}\check{A}^{N}_{\nu}.
  \end{equation}

   Obviously,
 (\ref{kin3}) and (\ref{kin4}) are not yet of the same form. In fact, there
are two important differences:
 \begin{enumerate}
 \item Eq.  (\ref{kin3}) is expressed in terms of the period matrix $\mathcal{N}_{AB}$ that corresponds to the symplectic basis  $(X^{A}, F_{B})$.
Eq. (\ref{kin4}), on the other hand, is expressed in terms of the period matrix $\check{\mathcal{N}}_{AB}$ that corresponds to the symplectic section
 $(\check{X}^{A},\check{F}_{B})$.
 \item  Both  (\ref{kin3}) and (\ref{kin4}) are already  expressed in terms
of $\check{A}_{\mu}^{I}$ and $\check{A}_{\mu}^{M}$. However, (\ref{kin3})
is still expressed in terms of $A_{\mu}^{0}$, whereas (\ref{kin4})
already contains the dual vector field $\check{A}_{\mu}^{0}$.                   Furthermore, (\ref{kin3}) contains a
curious  term    proportional to   (the last term in (\ref{kin3}))
\begin{equation}
\frac{e^{-1} g}{4}\epsilon^{\mu\nu\rho\sigma}F_{\mu\nu}^{0}\check{A}_{\rho}^{M}\check{A}_{\sigma}^{N} . \label{newcoupling}
\end{equation}
Such terms have been analyzed in the literature before \cite{dWLVP,dWHR}    (see also the more recent paper \cite{AFL}). In our case,
this term corresponds to some of    the standard gauged supergravity terms in
(\ref{kin4}) upon the  dualization of $\check{F}_{\mu\nu}^{0}  \leftrightarrow  F_{\mu\nu}^{0}$, as we will show in a moment.
  \end{enumerate}

We will now show the equivalence of (\ref{kin3}) and (\ref{kin4}) by transforming (\ref{kin4}) into (\ref{kin3}). As we have already mentioned, $\check{A}_{\mu}^{0}$ appears in (\ref{kin4}) only via its (Abelian)   curl
  $\check{F}_{\mu\nu}^{0}$ as it gauges the central charge.
In (\ref{kin4}),    $\check{A}_{\mu}^{0}$ can  therefore   be dualized to another  vector field
$C_{\mu}$ with Abelian field strength $C_{\mu\nu}$. As usual, this
is done by adding
 \begin{equation}
-\frac{e^{-1}}{4} \epsilon^{\mu\nu\rho\sigma}\check{F}_{\mu\nu}^{0}C_{\rho\sigma} =  \textrm{Im}[\check{F}_{ \mu\nu}^{ 0 + }  C^{\mu\nu + } ]
 \end{equation}
to the Lagrangian (\ref{kin4}).
 Varying  with respect to $C_{\mu\nu}^{ + }   $
and reinserting the resulting equation for $\check{F}_{\mu\nu}^{0 +  }$
gives
\begin{eqnarray}
e^{-1}  \check{\mathcal{L}}^{\textrm{vec}}_{\textrm{kin, dual}} & = &
\frac{1}{2} \textrm{Im} \Big[ -2 C_{\mu\nu}^{ + } Z^{\mu\nu  +  }
   +   \Big( \check{\mathcal{N}}_{\tI\tJ} -
 \frac{   \check{\mathcal{N}}_{0\tI}     \check{\mathcal{N}}_{0\tJ}}{\check{\mathcal{N}}_{00}} \Big)
 \check{\mathcal{F}}_{\mu\nu}^{\tI + } \check{\mathcal{F}}^{\mu\nu \tJ + }\nonumber \\
 & & \qquad -2 \frac{\check{\mathcal{N}}_{0\tI}}{\check{\mathcal{N}}_{00}}
 \check{\mathcal{F}}_{\mu\nu}^{\tI + } C^{\mu\nu + } -\frac{1}{\check{\mathcal{N}}_{00}} C_{\mu\nu}^{ + } C^{\mu\nu + } \Big]\Big|_{A^M=0} . \label{redlag6}
\end{eqnarray}
 In order to bring this to the form (\ref{kin3}), it remains
  to reexpress the $\check{\mathcal{N}}_{AB}$ in terms of the
 $\mathcal{N}_{AB}$ that appear in (\ref{kin3}). To this end, recall that
 the basis $(\check{X}^{A},\check{F}_{B})$ is essentially obtained from the basis
 $(X^{A},F_{B})$ by exchange of  $X^{0}$ with $F_{0}$ and $X^{M}$ with $F_{M}$
 (in fact with $D^{MN}F_{N}$). This exchange is implemented by the symplectic transformation matrix $\mathcal{S}$ of eq. (\ref{Sdef}):
 \begin{equation}
  \left( \begin{array}{c}
  \check{X}^{A} \\
  \check{F}_{B}
  \end{array} \right) = \mathcal{S} \left( \begin{array}{c}
  X^{A} \\
  F_{B}
  \end{array} \right) .
 \end{equation}
 It is convenient to decompose this transformation into two steps.
 In the first step, $X^{M}$ and $D^{MN}F_{N}$ are exchanged
 by multiplication with the symplectic matrix
 \begin{equation}
 \mathcal{S}_{1} =  \left(  \begin{array}{cccccc}
1 & 0 & 0 & 0 & 0 & 0 \\
0 & \delta_{I}^{J} & 0 &0& 0& 0\\
0 & 0 & 0 & 0 & 0 & D^{MN}\\
0 & 0 & 0 & 1 & 0 & 0 \\
0 & 0& 0& 0& \delta_{I}^{J}& 0 \\
0 & 0 & D_{MN}  &0&0&0
\end{array}
\right) .
 \end{equation}
 We call the resulting symplectic vector $(\tilde{X}^{A}, \tilde{F}_{B})$, i.e.
 \begin{equation}
  \left( \begin{array}{c}
  \tilde{X}^{A} \\
  \tilde{F}_{B}
  \end{array} \right) = \mathcal{S}_{1} \left( \begin{array}{c}
  X^{A} \\
  F_{B}
  \end{array} \right) .
 \end{equation}
 In a second step, $X^{0}$ and $F_{0}$ (which are now called $\tilde{X}^{0}$
and $ \tilde{F}_{0}$) are rotated by subsequent multiplication with the symplectic matrix
\begin{equation}
 \mathcal{S}_{2} =  \left(  \begin{array}{cccccc}
0 & 0 & 0 & 1 & 0 & 0 \\
0 & \delta_{I}^{J} & 0 &0& 0& 0\\
0 & 0 & \delta_{M}^{N} & 0 & 0 & 0\\
-1 & 0 & 0 & 0 & 0 & 0 \\
0 & 0& 0& 0& \delta_{I}^{J}& 0 \\
0 & 0 & 0  &0&0& \delta_{M}^{N}
\end{array}
\right) .
 \end{equation}
 Obviously,
 \begin{equation}
 \mathcal{S}   = \mathcal{S}_{2}\mathcal{S}_{1}
 \end{equation}
 and
 \begin{equation}
  \left( \begin{array}{c}
  \check{X}^{A} \\
  \check{F}_{B}
  \end{array} \right) = \mathcal{S}_{2} \left( \begin{array}{c}
  \tilde{X}^{A} \\
  \tilde{F}_{B}
  \end{array} \right) .
 \end{equation}
 The period matrix is likewise  computed in a two step process. First,
following eq. (\ref{Ntrafo}),  we determine
 \begin{equation}
 \tilde{\mathcal{N}} = (C_{1} + D_{1}\mathcal{N} ) (A_{1} + B_{1}\mathcal{N})^{-1}
 \end{equation}
 where
 \begin{equation}
\mathcal{S}_{1}=    \left(  \begin{array}{cc}
   A_{1} & B_{1} \\
   C_{1} & D_{1}
   \end{array}  \right)  .
  \end{equation}
 The result is
 \begin{equation}
 \tilde{\mathcal{N}}_{AB} =      \left(
 \begin{array}{ccc}
 \mathcal{N}_{00} - \mathcal{N}_{0M}\mathcal{N}^{MN}\mathcal{N}_{N0}  & \quad \mathcal{N}_{0I} - \mathcal{N}_{0M}\mathcal{N}^{MN}\mathcal{N}_{NI}
 & \quad \mathcal{N}_{0M}\mathcal{N}^{MN}D_{NP}\\
 \mathcal{N}_{I0} - \mathcal{N}_{IM}\mathcal{N}^{MN}\mathcal{N}_{N0}
 &  \quad \mathcal{N}_{IJ} - \mathcal{N}_{IM}\mathcal{N}^{MN}\mathcal{N}_{NJ}
 & \quad \mathcal{N}_{IM}\mathcal{N}^{MN}D_{NP}\\
 -D_{MN}\mathcal{N}^{NP}\mathcal{N}_{P0}
 & \quad -D_{MN}\mathcal{N}^{NP}\mathcal{N}_{PI}
 &  \quad C_{MN}\mathcal{N}^{NP}  D_{PR}\end{array}  \right)  \label{tildeN}
 \end{equation}
$\check{\mathcal{N}}_{AB}$ can then be obtained from
\begin{equation}
  \check{\mathcal{N}} = (C_{2} + D_{2}\tilde{\mathcal{N}} ) (A_{2} + B_{2}\tilde{\mathcal{N}})^{-1} ,
 \end{equation}
 where
 \begin{equation}
\mathcal{S}_{2}=    \left(  \begin{array}{cc}
   A_{2} & B_{2} \\
   C_{2} & D_{2}
   \end{array}  \right)  .
  \end{equation}
The result is
\begin{equation}
\check{\mathcal{N}}_{AB}= \frac{1}{\tilde{\mathcal{N}}_{00}}  \left( \begin{array}{cc}
-1 & \quad \tilde{\mathcal{N}}_{0\tI}\\
\tilde{\mathcal{N}}_{\tI 0} & \quad  (\tilde{\mathcal{N}}_{00} \tilde{\mathcal{N}}_{\tI\tJ} -\tilde{\mathcal{N}}_{0\tI} \tilde{\mathcal{N}}_{0\tJ})
\end{array}
\right)  . \label{checkN}
\end{equation}
We are now ready to show the equivalence of (\ref{kin3}) and (\ref{redlag6}).
First, using (\ref{checkN}), one rewrites (\ref{redlag6}) as
\begin{eqnarray}
e^{-1}  \check{\mathcal{L}}^{\textrm{vec}}_{\textrm{kin, dual}} & = &
\frac{1}{2} \textrm{Im} \Big[ -2 C_{\mu\nu}^{ + } Z^{\mu\nu  +  }
   +    \tilde{\mathcal{N}}_{\tI\tJ}
 \check{\mathcal{F}}_{\mu\nu}^{\tI + } \check{\mathcal{F}}^{\mu\nu \tJ +}\nonumber \\
 & & \qquad +2 \tilde{\mathcal{N}}_{0\tI}
 \check{\mathcal{F}}_{\mu\nu}^{\tI + } C^{\mu\nu + } +\tilde{\mathcal{N}}_{00} C_{\mu\nu}^{ + } C^{\mu\nu + } \Big] \Big|_{A^M=0} .
\end{eqnarray}
Using (\ref{tildeN}), and identifying
\begin{equation}
F_{\mu\nu}^{0}=C_{\mu\nu}  ,
\end{equation}
this becomes eq. (\ref{kin3}).

What we have thus shown, is that after the tensor fields are
eliminated, the theory is dual to a standard gauged supergravity
theory with gauge group  $K \ltimes \mathcal{H}^{n_T+1}$.  In order
to gauge this group in the standard way, its action has to be made
block diagonal on the symplectic section prior to the gauging. This
is done by going to a new symplectic basis
$(\check{X}^{A},\check{F}_{B})$, which is obtained from the
``natural'' basis $(X^{A},F_{B})$ by exchanging $X^{0}$ with $F_{0}$
and $X^{M}$ with $D^{MN}F_{N}$ by means of a symplectic rotation.
The same rotations have to be applied to the corresponding field
strengths $(F_{\mu\nu}^{0},G_{\mu\nu 0})$ and
$(F_{\mu\nu}^{M},G_{\mu\nu N})$, where they correspond to
electromagnetic duality transformations. After this transformation,
the gauging can be carried out in the standard way. In order to
recover the compactified theory with the tensor fields eliminated,
one finally has to re-dualize $\check{F}_{\mu\nu}^{0}$ \emph{after}
the gauging. This dualization essentially takes back the exchange of
$X^{0}$ with $F_{0}$ (and the corresponding exchange of
$F_{\mu\nu}^{0}$ and $G_{\mu\nu  0  }$), but leaves some unusual new
couplings of the form (\ref{newcoupling}). The vector fields
$B_{\mu}^{M}$ that descend from the 5D tensor fields are interpreted
as massive vector fields that gained their mass from eating the
scalars $A^{M}$, which disappeared from the action. The
$B_{\mu}^{M}$ are essentially   the magnetic duals of the
$A_{\mu}^{M}$ of the ungauged theory. This makes sense, as the 5D
tensors         $\hat{B}_{\hat{\mu}\hat{\nu}}^{M}$
from which the $B_{\mu}^{M}$ descend,    are also the duals of the 5D vector fields
$\hat{A}_{\hmu}^{M}$, from which the $A_{\mu}^{M}$ descend.


\subsection{The case of not completely reducible representations}
In this subsection, we briefly comment on the dimensionally reduced
theory corresponding to case (ii) of Section 3, where the decomposition
of the $(\tilde{n}+1)$-dimensional representation of $G$ with respect to the
the prospective 5D gauge group $K$ is reducible, but  not \emph{completely}
reducible. This case has been studied  in  ref.  \cite{5Dgroup}. In that case,
the vector fields $\hat{A}_{\hmu}^{I}$ still transform in the adjoint
representation of $K\subset G$ and have the standard  field strengths
$\hat{\mathcal{F}}_{\hmu\hnu}^{I}\equiv 2\partial_{[\hmu} \hat{A}_{\hnu]}^{I}
+gf_{JK}^{I}\hat{A}_{\hmu}^{J}\hat{A}_{\hnu}^{K}$. In addition to the
 transformation matrix $\Lambda_{IM}^{N}$ that acts only on the tensor fields
$\hat{B}_{\hmu\hnu}^{M}$, however, there is now also a transformation matrix
of the type $\Lambda_{IJ}^{M}$    \footnote{These matrices are called
$t_{IJ}{}^{M}$ in \cite{5Dgroup}.} that can mix the tensor fields with the field
strengths of the vector fields, so that the representation of $K$ is no longer
 block diagonal, i.e. completely reducible.

 This new matrix is related to a new allowed
set of components of the $C_{\ti\tj\tk}$ tensor, namely the components
of the form $C_{IJM}$ (which have to vanish in the completely reducible case
  (i) of Section 3):
\begin{equation}
C_{IJM}=-\sqrt{6}\Lambda_{(IJ)}^{N}\Omega_{NM} \ .
\end{equation}

The modifications that are necessary to perform such a gauging in a supersymmetric way are  the same as in the completely reducible case,
except for the following differences:

\begin{itemize}
\item The covariant derivative (\ref{covB}) of the tensor fields, $\hat{\mathcal{D}}_{[\hmu}
\hat{B}_{\hnu\hrho]}^{M}$, in the $\hat{B}^{N}\wedge \hat{\mathcal{D}}\hat{B}^{M}$
term of the five-dimensional Lagrangian (\ref{start}) gets an additional contribution due to the mixing matrix $\Lambda_{IJ}^{M}$:
\begin{equation}
\hat{\mathcal{D}}_{[\hat{\mu}}\hat{B}_{\hnu\hrho]}^M \rightarrow \partial_{[\hmu}
  \hat{B}_{\hnu\hrho]}^M +2g\Lambda_{IJ}^{M}\hat{A}_{[\hmu}^{I}\hat{\mathcal{F}}_{\hnu\hrho]}^{J} +g\hat{A}_{[\hmu}^{I}\Lambda_{IN}^{M}\hat{B}_{\hnu\hrho]}^{N}.
\end{equation}
\item There are new  Chern-Simons terms of the type $AAAF$ and $AAAAA$ beyond the already existing ones that are already displayed in  (\ref{start}) for the
completely reducible case:
\begin{equation}
\mathcal{L}_{\textrm{C.-S.}}^{\textrm{additional}}= -\frac{1}{2}\epsilon^{\hmu\hnu\hlambda\hrho\hsigma}\Omega_{MN}\Lambda_{IK}^{M}\Lambda_{FG}^{N} \hat{A}_{\hmu}^{I}\hat{A}_{\hnu}^{F}\hat{A}_{\hlambda}^{G}
 \left( -\frac{1}{2} g \hat{\mathcal{F}}_{\hrho\hsigma}^{K} + \frac{1}{10}g^{2}
 f_{HL}^{K}\hat{A}_{\hrho}^{H}\hat{A}_{\hsigma}^{L}  \right)  .
\end{equation}
 \item The new couplings enter the Killing vectors (and hence the covariant derivatives of the scalars) and the scalar potential via an implicit dependence
on $\Lambda_{IJ}^{M}$.
\end{itemize}

These modifications all have their counterparts in the dimensionally reduced theory in four dimensions,  and it is   straightforward to determine them from    an obvious   generalization of the equations displayed in the appendix.   One important aspect of the 4D theory,   however, can best be seen from  the way    the non-vanishing  $C_{IJM}$ terms influence the
transformation laws of the symplectic section under the translation of the  Kaluza-Klein scalars $A^{M}$  by $b^{M}$. Indeed,  the $F_{A}$ components of the symplectic section now have additional contributions from the new $C_{IJM}$ terms in the prepotential $F$, so
 we now have under
infinitesimal translation $z^{M} \rightarrow z^{M} + b^{M}$,
\begin{equation}
\left( \begin{array}{c}
X^{0}\\
X^{I}\\
X^{M}\\
F_{0}\\
F_{I}\\
F_{M}
\end{array} \right) \rightarrow
\left( \begin{array}{c}
X^{0}\\
X^{I}\\
X^{M}\\
F_{0}\\
F_{I}\\
F_{M}
\end{array} \right) + \left( \begin{array}{c}
0\\
0\\
b^M X^{0} \\
 -b^M F_{M}  \\
-       2    \sqrt{2}b^{M}C_{MNI}X^{N}  -2 \sqrt{2}b^{M}C_{IJM}X^{J} \\
-       2    \sqrt{2}b^{M}C_{MIN}X^{I}
\end{array} \right) .
\end{equation}
  We observe that, just as for the  completely reducible case,
  the  components $(F_{0},X^{I},F_{M})$ still transform among themselves.
  However, this is no longer true for the components $(X^{0},F_{I},X^{M})$
   if $C_{IJM}\neq0$  --- a clear difference  to  the
  completely reducible case with $C_{IJM}=0$.   In fact, the minimal set of components that contains the $F_{I}$ and closes under translations
  is in general $(X^{0},F_{I},X^{M},X^{I})$, which is too big for
  one half of a symplectic section. One might wonder whether
   perhaps some   linear combination of the $X^{M}$ and $X^{I}$ could be used instead of all the $X^{M}$ and $X^{I}$  in this set, so as to make the number of independent components smaller, but this would require a symplectic rotation that somehow trades the $F_{M}$ with that linear combination of the
   $X^{M}$ and $X^{I}$, just as we traded $F_{M}$ and $X^{M}$ using the matrix $\mathcal{S}_1$ in the completely reducible case. However, whereas $\mathcal{S}_{1}$ contained only Kronecker deltas and the matrix $D_{MN} \sim \Omega_{MN} $, i.e. a natural object of the 5D theory, there is no natural
   object with the index structure $\{\cdot\}_{M}^{I}$ that one can build
   from the 5D objects $\Omega_{MN},f_{IJ}^{K},\Lambda_{IM}^{N},\Lambda_{IJ}^{M},C_{IJK}$,
   which determine the whole theory.
  Thus, due
to the presence of  $C_{IJM}$ terms, it  seems in general not possible
to find a symplectic matrix  $\mathcal{S}$ that brings the gauge
transformations to block diagonal form. As a result, these theories in 4D  should, apart from some possible special cases,  involve
topological terms of the form studied in \cite{dWLVP,dWHR} in addition to the
standard Yang-Mills gauging, even after dualizations of the type discussed
in the previous subsections are performed.



\section{$CSO^*(2N)$ Gauged Supergravity Theories from Reduction of 5D Theories and Unified YMESGTs in Four Dimensions}

Unified 5D MESGTs are defined as those theories whose Lagrangian
admit a simple symmetry group under which all the vector fields,
including the ``graviphoton'', transform irreducibly. Among those 5D
MESGTs whose scalar manifolds are homogeneous spaces only four are
unified \cite{GZ4}. They are defined by the four simple
Euclidean Jordan algebras  of degree three,  $J_3^{\mathbb{A}}$, of $(3\times 3)$
Hermitian matrices over the four division algebras
$\mathbb{A}=\mathbb{R},\mathbb{C},\mathbb{H},\mathbb{O}$ \cite{GST1},
and their scalar manifolds are   actually   symmetric spaces, which we list below
:
\begin{eqnarray}
\mathcal{M}&=& SL(3,\mathbb{R})/ SO(3)\qquad (\tn=5)\nonumber\cr
\mathcal{M}&=& SL(3,\mathbb{C})/ SU(3)\qquad (\tn=8)\nonumber\cr
\mathcal{M}&=& SU^{*}(6)/ Usp(6)\qquad
(\tn=14)\nonumber\\
\mathcal{M}    &=& E_{6(-26)}/ F_{4}\qquad \qquad
(\tn=26)\label{magicalfamily},
\end{eqnarray}
where we have indicated the number of vector multiplets, $\tn$, for
each of these theories. In these cases, the symmetry groups $G$ of
these theories are simply the isometry groups $SL(3,\mathbb{R})$,
$SL(3,\mathbb{C})$, $SU^{\ast}(6)$ and $E_{6(-26)}$, respectively,
under which  the, respectively, $6$, $9$, $15$ and $27$ vector
fields $A_{\mu}^{\ti}$ transform irreducibly \cite{GST1}. Thus,
according to our definition, all of these four theories are unified
MESGTs. These supergravity theories are referred to as the magical
supergravity theories \cite{GST3} , because of their deep connection
with the Magic Square of Freudenthal, Rozenfeld and Tits \cite{FRT}.
Of these four unified MESGTs in five dimensions only the theory
defined by $J_3^{\mathbb{H}}$ can be gauged so as to obtain a
unified YMESGT    \footnote{In unified YMESGTs  all the vector
fields, including the graviphoton, transform in the adjoint
representation of a simple gauge group.} with the gauge group
$SO^*(6) \simeq SU(3,1)$.

As was shown in \cite{GZ4}, if one relaxes the requirement that the
scalar manifolds be homogeneous spaces, one finds three infinite
families of unified MESGTs in five dimensions. They are defined by
Lorentzian Jordan algebras of arbitrary degree over the four
associative division algebras $\mathbb{R},\mathbb{C},\mathbb{H}$.
These  Lorentzian  Jordan algebras $J_{(1,N)}^{\mathbb{A}} $ of degree
 $p  =N+1$ are realized by $(N+1)\times (N+1)$ matrices over $\mathbb{A}$ which are
 hermitian with respect to the Lorentzian metric  $\eta = (-,+,+,...,+)$:
\begin{equation}\label{eta}
(\eta X)^\dag = \eta X  \hspace{1cm} \forall  X\in
J_{(1,N)}^{\mathbb{A}} \ .
\end{equation}
  A general element, $U$, of $J_{(1,N)}^{\mathbb{A}} $ can be written in
the form
\begin{equation}
U=\left(\begin{array}{cc}
x & -Y^{\dagger}\\
Y & Z
\end{array}\right),
\end{equation}
where $Z$ is an element of the Euclidean subalgebra
$J_{N}^{\mathbb{A}} $ (i.e., it is a Hermitian $(N\times N)$-matrix
over $\mathbb{A}$), $x\in\mathbb{R}$, and $Y$ denotes an
$N$-dimensional  column vector over $\mathbb{A}$. Under their
(non-compact)   automorphism group,
$\textrm{Aut}(J_{(1,N)}^{\mathbb{A}})$,
   these  simple Jordan algebras $J_{(1,N)}^{\mathbb{A}}$ decompose into an
irreducible representation formed by the traceless elements plus a
singlet, which is given by the identity element of
$J_{(1,N)}^{\mathbb{A}}$ (i.e., by the unit matrix $\mathbf{1}$):
\begin{equation}
J_{(1,N)}^{\mathbb{A}}= \mathbf{1}\oplus \{ \textrm{traceless
elements}\} \equiv  \mathbf{1} \oplus J_{(1,N)_0}^{\mathbb{A}}  .
\end{equation}

By identifying  the structure constants ($d$-symbols)  of the
traceless elements of a Lorentzian Jordan algebra
$J_{(1,N)}^{\mathbb{A}}$ with the $C_{\ti\tj\tk}$ of a MESGT:
$C_{\ti\tj\tk}=d_{\ti\tj\tk}$, one obtains a unified MESGT, in which
all the vector fields transform irreducibly under the simple
automorphism group $\textrm{Aut}(J_{(1,N)}^{\mathbb{A}})$ of that
Jordan algebra. For $\mathbb{A}=\mathbb{R},\mathbb{C},\mathbb{H}$
one obtains in this way three infinite families of physically
acceptable unified MESGTs (one for each $N\geq2$).

In table 1  below, we list  all the simple Lorentzian Jordan algebras
of type $J_{(1,N)}^{\mathbb{A}} $, their automorphism groups, and
the numbers of vector and scalar fields in the unified 5D MESGTs
defined by them.

 \begin{table}[ht]

\begin{center}
\begin{displaymath}
\begin{array}{|c|c|c|c|c|}
\hline
~&~&~&~&~\\
J & D&  \textrm{Aut}(J)
& \textrm{No. of vector fields} & \textrm{No.  of scalars}  \\
\hline
~&~&~&~&~\\
J_{(1,N)}^{\mathbb R}& \frac{1}{2} (N+1)(N+2) & SO(N,1) &
\frac{1}{2}N(N+3) &\frac{1}{2}N(N+3)-1 \\
~&~&~&~&~\\
J_{(1,N)}^{\mathbb C} & (N+1)^2 & SU(N,1) & N(N+2) & N(N+2)-1 \\
~&~&~&~&~\\
J_{(1,N)}^{\mathbb H} &(N+1)(2N+1) & USp(2N,2) & N(2N+3) & N(2N+3)-1 \\
~&~&~&~&~\\
J_{(1,2)}^{\mathbb O} & 27& F_{4(-20)} & 26 & 25 \\
~&~&~&~&~ \\
\hline
\end{array}
\end{displaymath}
\end{center}

\caption{List of the simple Lorentzian Jordan algebras of type
$J_{(1,N)}^{\mathbb{A}}$.
 The columns show, respectively, their  dimensions $D$,
their automorphism groups $\textrm{Aut} (J_{(1,N)}^{\mathbb{A}})$,
the number of vector fields $(\tn+1)=(D-1)$ and the number of
scalars $\tn=(D-2)$ in the corresponding MESGTs.}
\end{table}

Note that the number of vector fields for the theories defined by
$J_{(1,3)}^{\mathbb R}$, $J_{(1,3)}^{\mathbb C}$ and
$J_{(1,3)}^{\mathbb H}$ are  9, 15 and 27, respectively. These are
exactly the same numbers of vector fields  as  in the magical theories
based on the norm forms of the Euclidean Jordan algebras
$J_{3}^{\mathbb C}$, $J_{3}^{\mathbb H}$ and $J_{3}^{\mathbb O}$,
respectively      (cf. eq. (\ref{magicalfamily})).   As was shown   in \cite{GZ4}, this is not an accident;
the magical MESGTs based on $J_{3}^{\mathbb C}$, $J_{3}^{\mathbb H}$
and $J_{3}^{\mathbb O}$ found in \cite{GST1} are \emph{equivalent}
(i.e. the cubic polynomials $\mathcal{V}(h)$ agree) to the ones
defined by the Lorentzian algebras $J_{(1,3)}^{\mathbb R}$,
$J_{(1,3)}^{\mathbb C}$ and $J_{(1,3)}^{\mathbb H}$, respectively.
This is a consequence of the fact that the generic norms of  the
degree 3 simple Euclidean Jordan algebras $J_{3}^{\mathbb C}$,
$J_{3}^{\mathbb H}$ and $J_{3}^{\mathbb O}$ coincide with the cubic
norms defined over  the traceless elements of degree four simple
Lorentzian Jordan algebras over $\mathbb{R}, \mathbb{C}$ and
$\mathbb{H}$ \cite{GZ4}. This implies that the only known unified
MESGT that is not covered by the     Table 1   is the magical theory
of \cite{GST1} based on the Euclidean Jordan algebra
$J_{3}^{\mathbb{R}}$ with $(\tn+1)=6$ vector fields and the target
space $\M=SL(3,\mathbb{R})/SO(3)$.
 Except for the theories defined by $J_{(1,3)}^{\mathbb R}$, $J_{(1,3)}^{\mathbb C}$ and
$J_{(1,3)}^{\mathbb H}$ the scalar manifolds of  MESGTs defined by
other simple Lorentzian Jordan algebras  are not homogeneous.

Of these three infinite families of unified MESGTs in five
dimensions only the family defined by $ J_{(1,N)}^{\mathbb{C}}$ can
be gauged so as to obtain an infinite family of unified YMESGTs with
the gauge groups $SU(N,1)$ \cite{GZ4}. As for the family defined by
the quaternionic Lorentzian Jordan algebras $
J_{(1,N)}^{\mathbb{H}}$, they can be gauged with the gauge groups
$SU(N,1)$ while dualizing the remaining $N(N+1)$ vector fields to
tensor fields in five dimensions.

 Let us now analyze the dimensional reduction of the 5D YMESGTs
with the gauge group $SU(N,1)$ coupled to $N(N+1)$ tensor fields.    From the results of section 6 it follows that the corresponding four
dimensional theory is  dual to a standard $\mathcal{N}=2$
YMESGT  with the gauge group $SU(N,1) \ltimes \mathcal{H}^{N(N+1)+1}
$. However, the group $SU(N,1) \ltimes \mathcal{H}^{N(N+1)+1}$ can
be obtained by contraction from the simple noncompact group
$SO^*(2N+2)$. This is best seen by considering the three graded
decomposition of the Lie algebra of $SO^*(2N+2)$ with respect to the
Lie algebra of its subgroup $SU (N,1) \times U(1) $
\[ \mathfrak{so}^*(2N+2) = \mathfrak{g}^{-1} \oplus [
\mathfrak{su}(N,1)\times \mathfrak{u}(1) ] \oplus \mathfrak{g}^{+1}
\] where grade $+1$ and $-1$ subspaces transform in the
antisymmetric tensor representation of $SU(N,1)$ and its conjugate,
respectively. By rescaling the generators belonging to the grade
$\pm 1$ spaces and redefining the generators in the limit in which  the    scale
parameter goes to infinity one obtains the Lie algebra isomorphic to
the Lie algebra of $SU(N,1) \ltimes \mathcal{H}^{N(N+1)+1}$.  Such
contractions arise in the pp-wave limits of spacetime groups and
were studied in \cite{ppwave}.    We shall denote the contracted
algebra as $CSO^*[2N+2\| U(N,1)]$. For $N=3$, $CSO^*[8\| U(3,1)]$
coincides with the contraction of $SO^*(8)$ denoted as $CSO^*(6,2)$
by Hull \cite{Hull}  since $SO^*(6)$ is isomorphic to $SU(3,1)$.

Now the MESGT theory defined by $J_{(1,3)}^{\mathbb H}$ can be
gauged directly in four dimensions so as to obtain a unified YMESGT
with the gauge group $SO^*(8)=SO(6,2)$. By contracting this unified
theory, one can obtain the $CSO^*[8\| U(3,1)]= CSO^*(6,2)$ gauging
directly in four dimensions, which is consistent with the above
observation.

 In \cite{GMcRZ1}     it
was pointed out that the three infinite families of 4D MESGTs
defined by Lorentzian Jordan algebras might  admit symplectic
sections in which all the vector fields transform irreducibly under
the reduced structure groups of the corresponding Jordan algebras.
Since their reduced structure groups are simple  they would be
unified MESGTs in four dimensions as well.  Of these three infinite
families of unified MESGTs only the family defined by the
quaternionic Jordan algebras $J_{(1,N)}^{\mathbb H}$  could  then be
gauged so as to obtain unified YMESGTs with gauge groups
$SO^*(2N+2)$ in four dimensions. The fact that the dimensional
reduction of the five dimensional
 gauged YMESGTs with gauge groups $SU(N,1)$ coupled to $N(N+1)$
tensor fields leads to contracted versions of the $SO^*(2N+2)$
gauged YMESGTs is evidence for the existence of this infinite family
of unified YMESGTs.

\section{Some comments on the scalar potential}
We already mentioned in the Introduction that    5D
     noncompact     YMESGTs  with    tensor multiplets and R-symmetry gauging
 provide the only known examples of  stable de Sitter ground states
 in higher-dimensional gauged supergravity theories \cite{GZ2,CS}.
   In this paper, we considered the dimensional reduction of
    5D YMESGTs with tensor fields (but   without R-symmetry gauging) to 4D and found that the resulting theories have non-Abelian noncompact gauge groups in 4D which are of the form $K\ltimes     \mathcal{H}^{n_{T}+1}$.
    Non-compact non-Abelian gauge groups were also found essential
    for stable de Sitter vacua in   4D,  $\mathcal{N}=2$ supergravity
    in  \cite{4dS}.
    Interestingly, the vectors that gauge      the Heisenberg algebra
    require a symplectic rotation relative to the vector fields that gauge the
    5D part $K$ of the 4D gauge group in order to  bring  the 4D gauging into
     the standard block diagonal form.  Apart from the semi-direct vs. direct structure, this is reminiscent of the      de Roo-Wagemans angles
     that were also  found    to be   important for stable de Sitter ground states in 4D, $\mathcal{N}=2$ supergravity  in \cite{4dS}.
     As a third ingredient for stable de Sitter vacua in  4D, $\mathcal{N}=2$ supergravity, the authors of   \cite{4dS}  identified gaugings of the R-symmetry group, which are also important in 5D  \cite{GZ2,CS}.

     One might now wonder whether these findings might perhaps have something to with each other.   Let us therefore take a look at the scalar potential
     of the   dimensional reduced  YMESGTs with tensor fields. From eq. (\ref{totalpot}), we have
     \begin{equation} \label{totalpot2}
P=e^{-\sigma}P^{(T)} (h^{\ti}) +\frac{3}{4} e^{-3\sigma}{\stackrel{\circ}{a}}_{\tI\tJ} (A^{I}M_{I\tK}^{\tI} h^{\tK})
(A^{J}M_{J\tilde{L}}^{\tJ}h^{\tilde{L}})
     \end{equation}
     where the first term is simply the dimensional reduction of the 5D
     scalar potential and the second term comes from the 5D kinetic term
     of the scalar fields. If we had instead started from a      5D, $\mathcal{N}=2$  YMESGT with tensor fields \emph{and} R-symmetry gauging,
     we would have gotten an additional term in 4D of the form
       \begin{equation}\label{Rpot}
       e^{-\sigma}P^{(R)}(h^{\ti})
       \end{equation}
     which is just   the dimensional reduction of the 5D scalar potential due to the R-symmetry gauging (the R-symmetry gauging does not affect the scalar fields, and therefore there is no analogue of the second term
     of eq. (\ref{totalpot2}) in addition to the already existing one.).
It is easy to convince oneself that the second term in (\ref{totalpot2})
is a positive definite real form for the $A^I$. A solution with $\langle A^{I} \rangle=0$ is therefore a solution without tachyonic directions in the $A^{I}$  space.  The first term in  (\ref{totalpot2}) and the term (\ref{Rpot})
 only depend on the $h^{\ti}$ and $\sigma$. Setting the $h^{\ti}$ equal to their values that are known to lead to a stable de Sitter vacuum in five dimensions
 in the models discussed in \cite{GZ2,CS} would then lead to a de Sitter
 point in 4D as well with the $h^{\ti}$ having positive masses. Unfortunately, however, this point would not be a critical point of the potential
 due to the runaway behaviour in the $\sigma$ direction. In order to fix $\sigma$ at finite values,
 one would have to allow the second term in (\ref{totalpot2}) to be non-zero.
 But then one would have to be at a point where    $\langle A^{I} \rangle\neq 0$,
 which might require other values of the $h^{\ti}$ that no longer correspond to
 the   stable de Sitter vacua that are known from five dimensions.

 A careful investigation of the scalar potential (\ref{totalpot2}) perhaps together with a gauging of the 4D R-symmetry group might lead to many
 interesting types of critical points, but is beyond the scope of this paper.

\appendix
\section{Appendix A}
This appendix lists the   the dimensional reductions of the
individual terms  of the Lagrangian (\ref{start}) using the
decompositions       (\ref{fuenfbein}),    (\ref{vector}) and (\ref{tensor}).
 \subsection{The Einstein-Hilbert term}
The  Einstein-Hilbert term  in (\ref{start})  leads to the same
four-dimensional terms as in the ungauged case,
\begin{eqnarray}
\mathcal{L}^{(5)}_{\textrm{E.-H.}} &\equiv& -\frac{1}{2}\he \hat{R} \Rightarrow \nonumber\\
 e^{-1} \mathcal{L}_{\textrm{E.-H.}}^{(4)}&=& -\frac{1}{2}R
 -\frac{1}{2}e^{3\sigma}W_{\mu\nu}W^{\mu\nu}-\frac{3}{4}\partial_{\mu}\sigma
 \partial^{\mu}\sigma
\end{eqnarray}

\subsection{The $\hat{\mathcal{H}}\hat{\mathcal{H}}$-term}
The $\hat{\mathcal{H}}\hat{\mathcal{H}}$-term in (\ref{start}) reduces to
\begin{eqnarray}
\mathcal{L}^{(5)}_{\hat{\mathcal{H}}\hat{\mathcal{H}}} & \equiv &
-\frac{1}{4}\he {\stackrel{\circ}{a}}_{\tI\tJ}
\hat{\mathcal{H}}_{\hat{\mu}\hat{\nu}}^{\tilde{I}}
\hat{\mathcal{H}}^{\tilde{J}\hat{\mu}\hat{\nu}} \Rightarrow
\nonumber\\
e^{-1} \mathcal{L}^{(4)}_{\hat{\mathcal{H}}\hat{\mathcal{H}}}&=&
-\frac{1}{4}e^{\sigma}{\stackrel{\circ}{a}}_{IJ}(\mathcal{F}_{\mu\nu}^{I}+2W_{\mu\nu}A^{I}
 )( \mathcal{F}^{J\mu\nu}+2W^{\mu\nu}A^{J}) \nonumber\\
 &&-\frac{1}{2}e^{\sigma}{\stackrel{\circ}{a}}_{IM}(\mathcal{F}_{\mu\nu}^{I}+2W_{\mu\nu}A^{I})B^{M\mu\nu}-\frac{1}{4}
e^{\sigma}{\stackrel{\circ}{a}}_{MN}B_{\mu\nu}^{M}B^{N\mu\nu} \nonumber\\
&&-\frac{1}{2}e^{-2\sigma}{\stackrel{\circ}{a}}_{IJ}(\mathcal{D}_{\mu} A^{I})(\mathcal{D}^{\mu}A^{J}) -e^{-2\sigma}{\stackrel{\circ}{a}}_{IM}(\mathcal{D}_{\mu}A^{I})B^{\mu M} \nonumber \\
&& -\frac{1}{2}e^{-2\sigma}{\stackrel{\circ}{a}}_{MN} B_{\mu}^{M}B^{\mu N} ,
\end{eqnarray}
where
\begin{eqnarray}
\mathcal{D}_{\mu}A^{I} & \equiv  & \partial_{\mu} A^{I} +g A_{\mu}^{J}f_{JK}^{I}A^{K}\\
\mathcal{F}_{\mu\nu}^{I}  &  \equiv  &    2\partial_{[\mu}A_{\nu]}^{I} + gf_{JK}^{I}A_{\mu}^{J}A_{\nu}^{K}.
\end{eqnarray}

\subsection{The scalar kinetic term}
Using (\ref{kin}) and
\begin{equation}
K_{I}^{\tx} (\partial_{\tx} h^{\tI})= M_{(I)   \tJ     }^{\tI} h^{\tJ},
\end{equation}
the 5D scalar kinetic term can be rewritten as
\begin{equation}
\mathcal{L}^{(5)}_{\textrm{scalar}} \equiv
-\frac{\hat{e}}{2}g_{\tx\ty}(\hat{\mathcal{D}}_{\hat{\mu}}\varphi^{\tx})
(\hat{\mathcal{D}}^{\hat{\mu}}\varphi^{\ty}) = -\frac{3 \hat{e}}{4}
{\stackrel{\circ}{a}}_{\tI\tJ}(\hat{\mathcal{D}}_{\hat{\mu}}h^{\tI})(\hat{\mathcal{D}}^{\hat{\mu}}
h^{\tJ}),
\end{equation}
where
\begin{equation}
\hat{\mathcal{D}}_{\hmu}h^{\tI} \equiv  \partial_{\hmu} h^{\tI} + g \hA_{\hmu}^{I}M_{I\tK}^{\tI}h^{\tK}.
\end{equation}
Upon dimensional reduction, this becomes
\begin{eqnarray}
\mathcal{L}^{(4)}_{\textrm{scalar}} &=& -\frac{3e}{4}
{\stackrel{\circ}{a}}_{\tI\tJ}(\mathcal{D}_{\mu}h^{\tI})(\mathcal{D}^{\mu}h^{\tJ})\nonumber\\
&& -\frac{3e}{4}g^2 e^{-3\sigma}{\stackrel{\circ}{a}}_{\tI\tJ}
(A^{I}M_{I\tK}^{\tI} h^{\tK})
(A^{J}M_{J\tilde{L}}^{\tJ}h^{\tilde{L}}),
\end{eqnarray}
where now the covariant derivative is with respect to the Kaluza-Klein invariant vector fields, $A_{\mu}^{I}$,
\begin{equation}
\mathcal{D}_{\mu}h^{\tI} \equiv  \partial_{\mu} h^{\tI} + g A_{\mu}^{I}M_{I\tK}^{\tI}h^{\tK}.
\end{equation}

\subsection{The Chern-Simons term}
The 5D Chern-simons term
\begin{eqnarray}
\hat{e}^{-1}  \mathcal{L}^{(5)}_{\textrm{C.S.}} & \equiv &
\frac{\hat{e}^{-1}}{6\sqrt{6}}C_{IJK}\hat{\epsilon}^{\hat{\mu}\hat{\nu}\hat{\rho}\hat{\sigma}\hat{\lambda}}\Big\{
\hat{F}_{\hat{\mu}\hat{\nu}}^{I}
{\hat{F}}_{\hat{\rho}\hat{\sigma}}^{J}
\hA_{\hat{\lambda}}^{K}+\frac{3}{2}g{\hat{F}}_{\hmu\hnu}^{I}\hA_{\hrho}^{J}
(f_{LF}^{K}\hA_{\hsigma}^{L}\hA_{\hlambda}^{F})\nonumber\\
&& \qquad \qquad \qquad
+\frac{3}{5}g^{2}(f_{GH}^{J}\hA_{\hnu}^G \hA_{\hrho}^{H})(f_{LF}^{K}\hA_{\hsigma}^{L}\hA_{\hlambda}^{F})\hA_{\hmu}^{I}
 \Big\}
\end{eqnarray}
reduces as follows
\begin{eqnarray}
e^{-1} \mathcal{L}_{\textrm{C.S.}}^{(4)}  &=& \frac{e^{-1}}{2\sqrt{6}} C_{IJK}\epsilon^{\mu\nu\rho\sigma} \Big\{
\mathcal{F}_{\mu\nu}^{I}\mathcal{F}_{\rho\sigma}^{J}A^{K} + 2 \mathcal{F}_{\mu\nu}^{I}W_{\rho\sigma}A^{J}A^{K}\nonumber\\ & & \qquad \qquad  \qquad  +\frac{4}{3}W_{\mu\nu}W_{\rho\sigma} A^{I}A^{J}A^{K} \Big\}
\end{eqnarray}

\subsection{The   $  \hat{B}  \hat{\mathcal{D}}  \hat{B}  $  term}
Using the decomposition (\ref{tensor}),  the 5D     $\hat{B}\hat{\mathcal{D}}\hat{B}$ term becomes
\begin{eqnarray}
\mathcal{L}^{(5)}_{\hat{B}\hat{\mathcal{D}}\hat{B}} &\equiv&
\frac{1}{4g}\hat{\epsilon}^{\hmu\hnu\hrho\hsigma\hlambda}\Omega_{MN}\hB_{\hmu\hnu}^{M}\hcD_{\hrho}\hB_{\hsigma\hlambda}^{N} \nonumber \\
 &=& \frac{1}{g} \epsilon^{\mu\nu\rho\sigma}\Omega_{MN}B_{\mu\nu}^{M}
(\partial_{\rho} B_{\sigma}^{N}+ g A_{\rho}^{I}\Lambda_{IP}^{N}B_{\sigma}^{P})\nonumber\\
&& +\frac{1}{g}
\epsilon^{\mu\nu\rho\sigma}\Omega_{MN}W_{\mu\nu}B_{\rho}^{M}B_{\sigma}^{N}
+\frac{1}{2\sqrt{6}}
C_{MNI}\epsilon^{\mu\nu\rho\sigma}B_{\mu\nu}^{M} B_{\rho\sigma}^{N}
A^{I}.
\end{eqnarray}

\subsection{The 5D scalar potential}
  The   5D   scalar potential term reduces to
  \begin{equation}
\mathcal{L}^{(5)}_{\textrm{pot}} \equiv  -\he g^2 P^{(T)} = -g^{2}e e^{-\sigma}P^{(T)}.
  \end{equation}

Putting everything together, and regrouping some terms,
one then arrives at   the dimensionally reduced YMESGT
with tensor fields written in eq. (\ref{redlag2}).

{\bf Acknowledgement:}          M.Z. would like to thank
G.~Dall'Agata, J.~Louis, H.~Samtleben and    S.~Vaul\`{a}   for
discussion and correspondence. M.G. would like to thank the
organizers of "Mathematical Structures in String Theory 2005"
workshop at KITP where part of this work was done. This work was
supported in part by the National Science Foundation under Grant No.
PHY-0245337 and Grant No. PHY99-07949. Any opinions, findings and
conclusions or recommendations expressed in this material  are those
of the author and do not necessarily reflect the views of  the
National Science Foundation. The work of M.Z. is supported    by the
German Research Foundation (DFG) within the Emmy-Noether-Program
         (ZA 279/1-2).

\end{document}